%% file: main.tex
\documentclass[10pt,journal,compsoc]{IEEEtran}

\ifCLASSOPTIONcompsoc
  \usepackage[nocompress]{cite}
\else
  \usepackage{cite}
\fi

\usepackage{graphicx}
\usepackage{scrtime}
\usepackage[dont-mess-around]{fnpct}
\usepackage{url}
\usepackage{epsfig}
\usepackage{amssymb, amsmath, mathtools}
\usepackage{empheq}
\usepackage{float}
\usepackage{subfloat}
\usepackage{caption}
\usepackage{array}
\usepackage{tabularx}
\usepackage{verbatim}
\usepackage{fancyvrb}
\usepackage{mathtools}
\usepackage{paralist}
\usepackage[ruled,linesnumbered,vlined]{algorithm2e}
\usepackage{cite}
\usepackage[tight,footnotesize]{subfigure}
\usepackage{float}
\usepackage{subfloat}
\usepackage{multirow}
\usepackage{transparent}
\usepackage{multicol}
\usepackage{esvect}
\usepackage{bm}
\usepackage{xfrac}
\usepackage{slashbox}
\usepackage{footnote}
\usepackage{soul,color}
\usepackage{todonotes}

\newtheorem{mydef}{Definition}

\newcommand{\spara}[1]{\smallskip\noindent{\bf{#1}}}
\newcommand{\smallSpacing}{\renewcommand{\baselinestretch}{0.9} \normalsize}

\begin{document}

\title{RELINE: Point-of-Interest Recommendations using Multiple Network Embeddings}

\author{Giannis~Christoforidis, Pavlos~Kefalas, 
    Apostolos~N.~Papadopoulos and Yannis~Manolopoulos
\IEEEcompsocitemizethanks{\IEEEcompsocthanksitem Giannis Christoforidis, Pavlos Kefalas and Apostolos N. Papadopoulos are with the Aristotle University of Thessaloniki, Greece. Yannis Manolopoulos is with the Open University of Cyprus. \protect\\
E-mail: {icchrist,kefalasp,papadopo}@csd.auth.gr, yannis.manolopoulos@ ouc.ac.cy
}
}


\IEEEtitleabstractindextext{%
\begin{abstract}
The rapid growth of users' involvement in Location-Based Social Networks (LBSNs) has led to the expeditious growth of the data on a global scale. The need of accessing and retrieving relevant information close to users' preferences is an open problem which continuously raises new challenges for recommendation systems. The exploitation of Points-of-Interest (POIs) recommendation by existing models is inadequate due to the sparsity and the cold start problems. To overcome these problems many models were proposed in the literature, but most of them ignore important factors such as: geographical proximity, social influence, or temporal and preference dynamics, which tackle their accuracy while personalize their recommendations. In this work, we investigate these problems and present a unified model that jointly learns user’s and POI dynamics. Our proposal is termed RELINE (\textbf{RE}commendations with mu\textbf{L}t\textbf{I}ple \textbf{N}etwork \textbf{E}mbeddings). More specifically, RELINE captures: $i$) the {\it social}, $ii$) the {\it geographical}, $iii$) the {\it temporal influence}, and $iv$) the {\it users' preference dynamics}, by embedding eight relational graphs into one shared latent space. We have evaluated our approach against state-of-the-art methods with three large real-world datasets in terms of accuracy. Additionally, we have examined the effectiveness of our approach against the cold-start problem. Performance evaluation results demonstrate that significant performance improvement is achieved in comparison to existing state-of-the-art methods.
\end{abstract}

}

\maketitle

\IEEEpeerreviewmaketitle

\input{intro.tex}
\input{related.tex}
\input{preliminaries.tex}
\input{proposed.tex}
\input{evaluation.tex}
\input{conclusions.tex}

\smallSpacing
\bibliographystyle{abbrv}
\bibliography{bibliography}  

\begin{IEEEbiography}[{\includegraphics[width=1in,height=1.3in,clip,keepaspectratio]{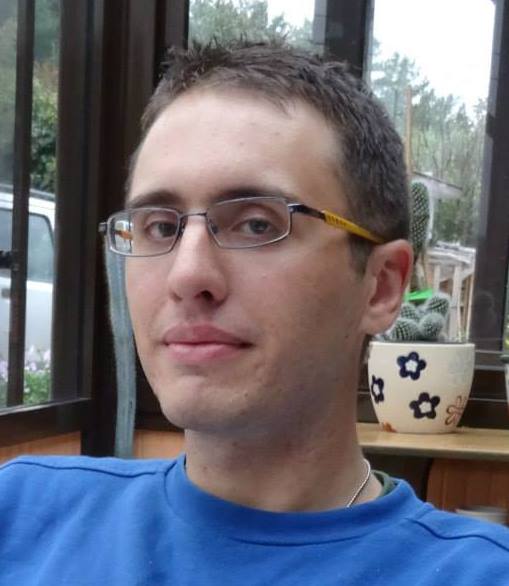}}]
{Giannis Christoforidis} received his 5-year Diploma degree in Informatics and Telecommunications Engineering from University of Western Macedonia (2014) and his M.Sc. degree in Networking and Web Development from Aristotle University of Thessaloniki (2018). Currently, he is a Ph.D. candidate at the same University and his research interests are based on knowledge discovery from big data.
\end{IEEEbiography}
\begin{IEEEbiography}[{\includegraphics[width=1in,height=1.25in,clip,keepaspectratio]{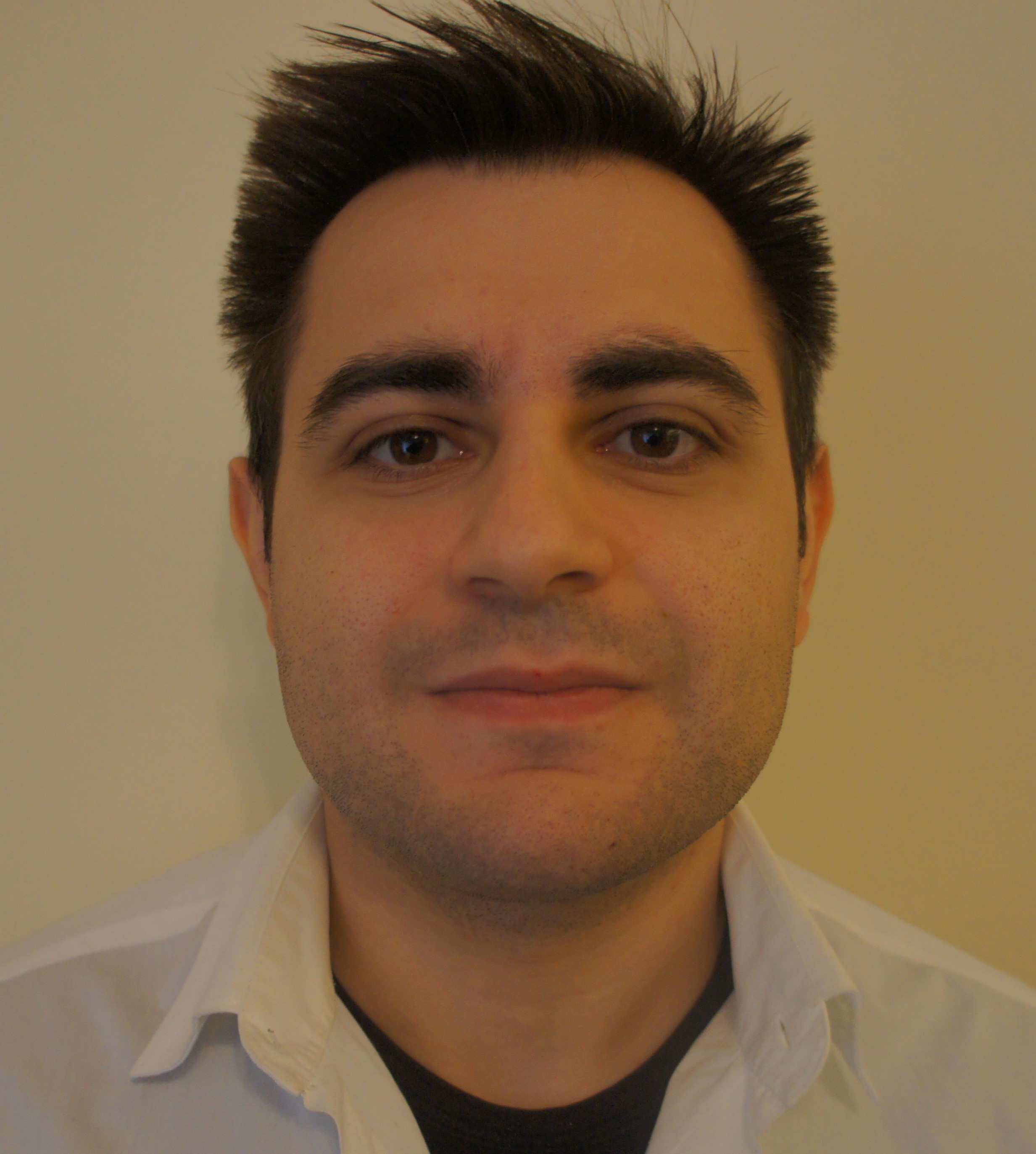}}]
{Pavlos Kefalas} received his B.Sc. degree in Computer Science (2009) from University of Crete, and his M.Sc. degree in Information Systems from Aristotle University of Thessaloniki (2011). He received his Ph.D. degree in recommendation models for dynamic spatio-temporal big data from the same university (2017). Currently, he is director of Artificial Intelligence and Machine Learning at Connex One Ltd. (UK). His research interests include recommender systems, reasoning, information retrieval, machine learning, distributed databases, spatio-temporal mining, and graph mining. 
\end{IEEEbiography}
\begin{IEEEbiography}[{\includegraphics[width=1in,height=1.25in,clip,keepaspectratio]{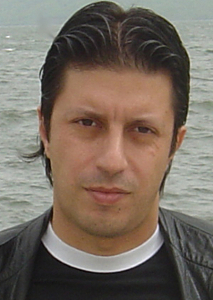}}]
{Apostolos N. Papadopoulos} received his 5-year Diploma degree in Computer Engineering and Informatics from University of Patras (1994) and his Ph.D. degree from Aristotle University of Thessaloniki (2000). His research interests include databases, data mining and big data analytics. The paper entitled ``SkyGraph: An Algorithm for Important Subgraph Discovery, received the award for the best Knowledge Discovery paper in ECML/PKDD 2008. Moreover, the paper ``Metric-Based Top-$k$ Dominating Queries'' that was presented in EDBT 2014, has been selected as the best paper and an extended version appears in ACM Trans. on Database Systems. Currently, he is an Associate Professor at the Department of Informatics of the Aristotle University of Thessaloniki. 
\end{IEEEbiography}
\begin{IEEEbiography}[{\includegraphics[width=1in,height=1.25in,clip,keepaspectratio]{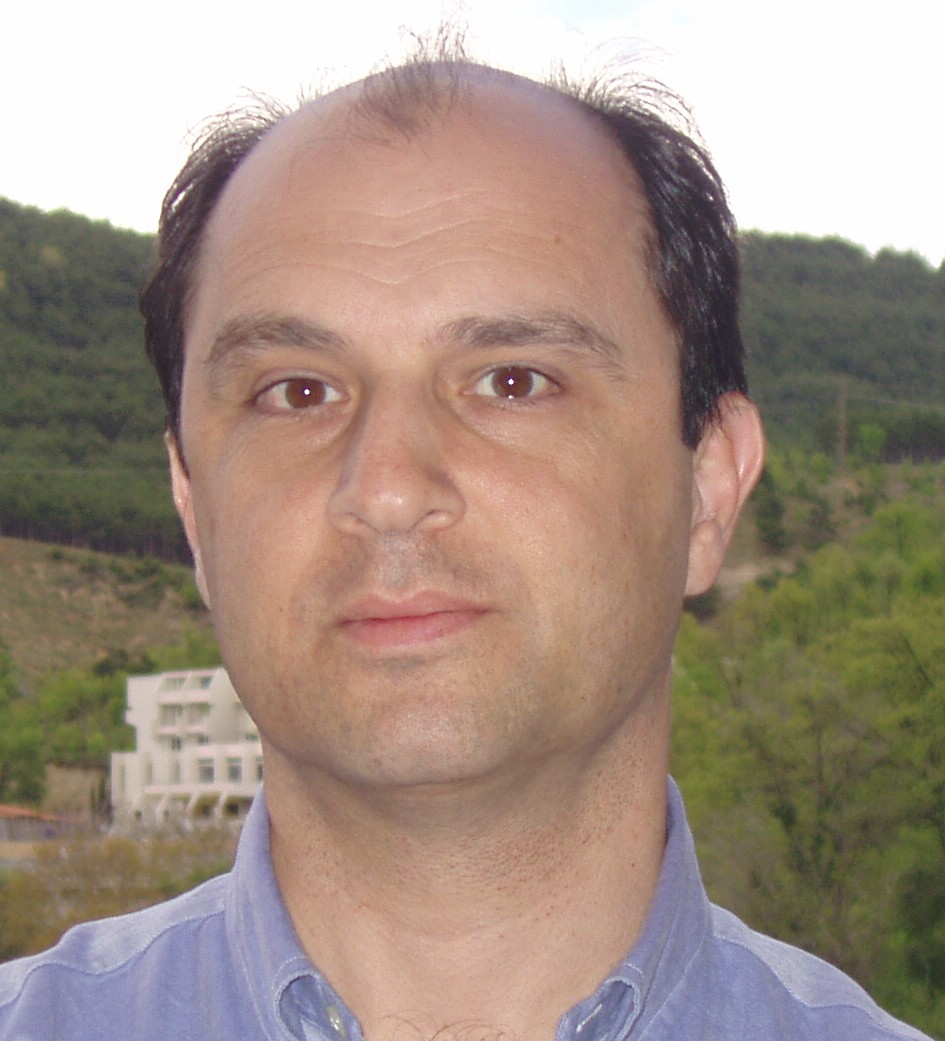}}]
{Yannis Manolopoulos} is a Professor and Vice-rector with the Open University of Cyprus. In the past, he has been with the University of Toronto, the University of Maryland at College Park, the University of Cyprus and the Aristotle University of Thessaloniki. He has served as Rector of the University of Western Macedonia, Greece and Vice-Chair of the Greek Computer Society. He has co-authored 5 monographs published by Springer, 8 textbooks in Greek, as well as $>$300 journal and conference papers related to Data Management. He received $>$13000 citations from $>$2000 distinct academic institutions and 5 best paper awards from SIGMOD, ECML/PKDD and MEDES (2) and ISSPIT conferences. He has also served as main co-organizer of several major fora, among others: ADBIS'2002, SSTD'2003, SSDBM'2004, ICEIS'2006, EANN'2007, ICANN'2010, AIAI'2012, WISE'2013, CAISE'2014, MEDI'2015, ICCCI' 2016, TPDL'2017, DAMDID'2017, DASFAA'2018, EAIS'2018, WIMS'2018, IDEAS'2019, MEDES'2019 conferences. He has acted as evaluator for funding agencies in Austria, Canada, Cyprus, Czech Republic, Estonia, EU, Georgia, Greece, Hong-Kong, Israel, Italy, Lithuania, Poland and Russia. Currently, he serves in the Editorial Board of Information Systems Journal, The World Wide Web Journal, The Computer Journal, among others. 
\end{IEEEbiography}

\end{document}

%% file: intro.tex
\section{Introduction}
\label{sec:Introduction}

\IEEEPARstart{R}{ecently}, \textit{Online Social Networks} (OSN) incorporated geographical information into their content which triggered new functionalities and introduced the concept of \textit{Location-based Social Networks} (LBSNs). In such networks, such as Facebook Places\footnote{\url{www.facebook.com/places}}, Foursquare\footnote{\url{www.foursquare.com}}, Yelp\footnote{\url{www.yelp.com}}, users may share their interests along with spatial dimension to obtain recommendations for possibly interesting places based on their recent history. Learning users' history is a crucial task for these models, to provide meaningful suggestions for \textit{Points-of-Interest} (POIs). Unfortunately, factors like sparsity, heterogeneity and multidimensionality pose significant challenges that increase the problem complexity with a large impact on effectiveness and efficiency. 

There has been extensive research on the topic, which primarily focused on users' relations with geographical information over user-location bipartite networks. However, such approaches failed to eliminate the sparsity problem since they miss preference dynamics and auxiliary information related to users' interactions. Similarly, other works enrich user-location information with social network ties but still miss all aforementioned factors. Moreover, research considering the temporal evolution of users' preferences still fail to deal with sparsity because there is additional contextual information related to users which changes along with their preferences over time. 

\begin{figure*}[!t]
\centering	
	\begin{subfigure}[Spatial Behavior]{
	\centering			
\includegraphics[height=3.5cm, width=4.2cm]{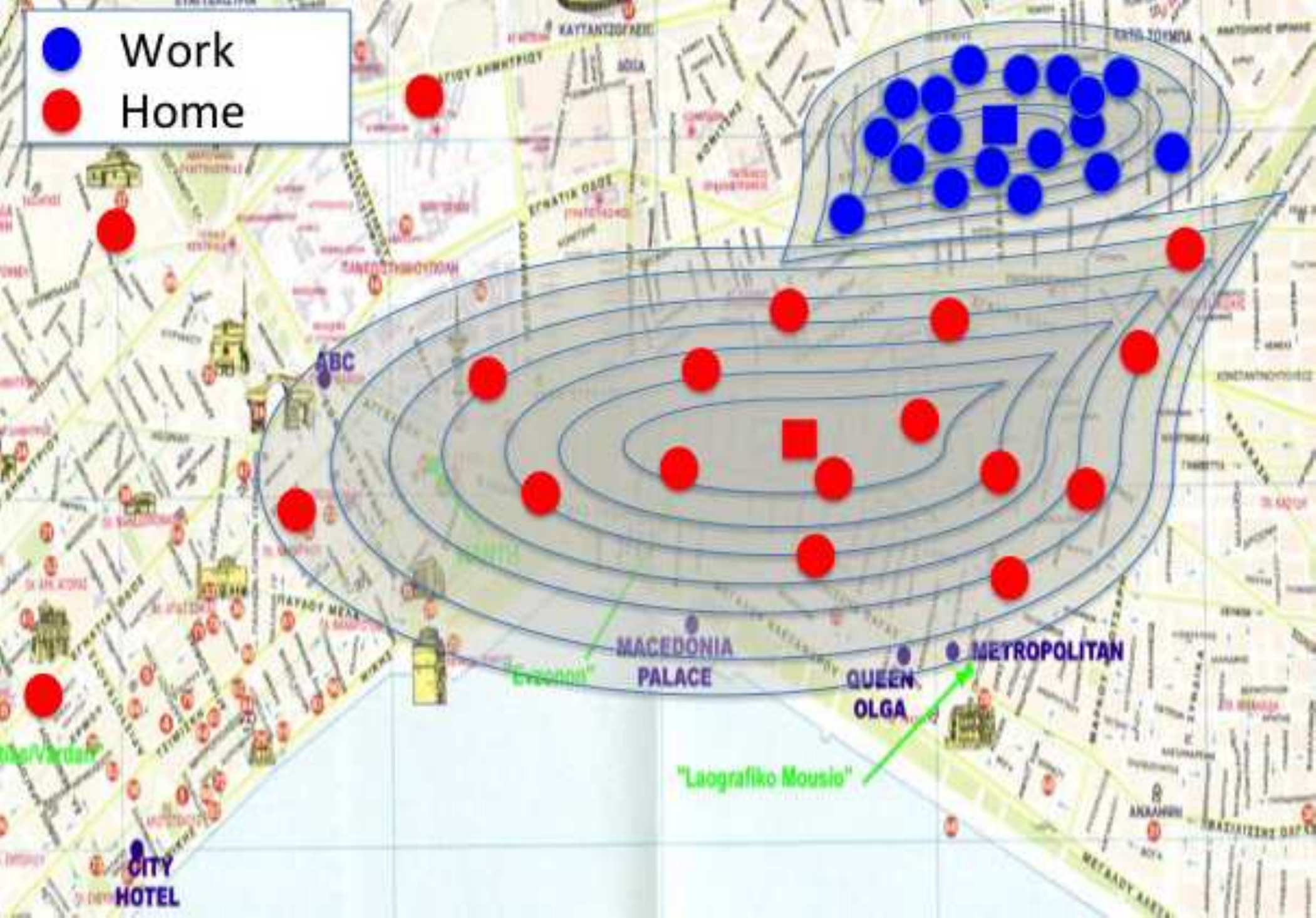}
		\label{fig:UserSpatialBehavior}
		}
	\end{subfigure}
	\begin{subfigure}[Temporal Behavior]{
	\centering	
\includegraphics[height=3.5cm, width=4.2cm]{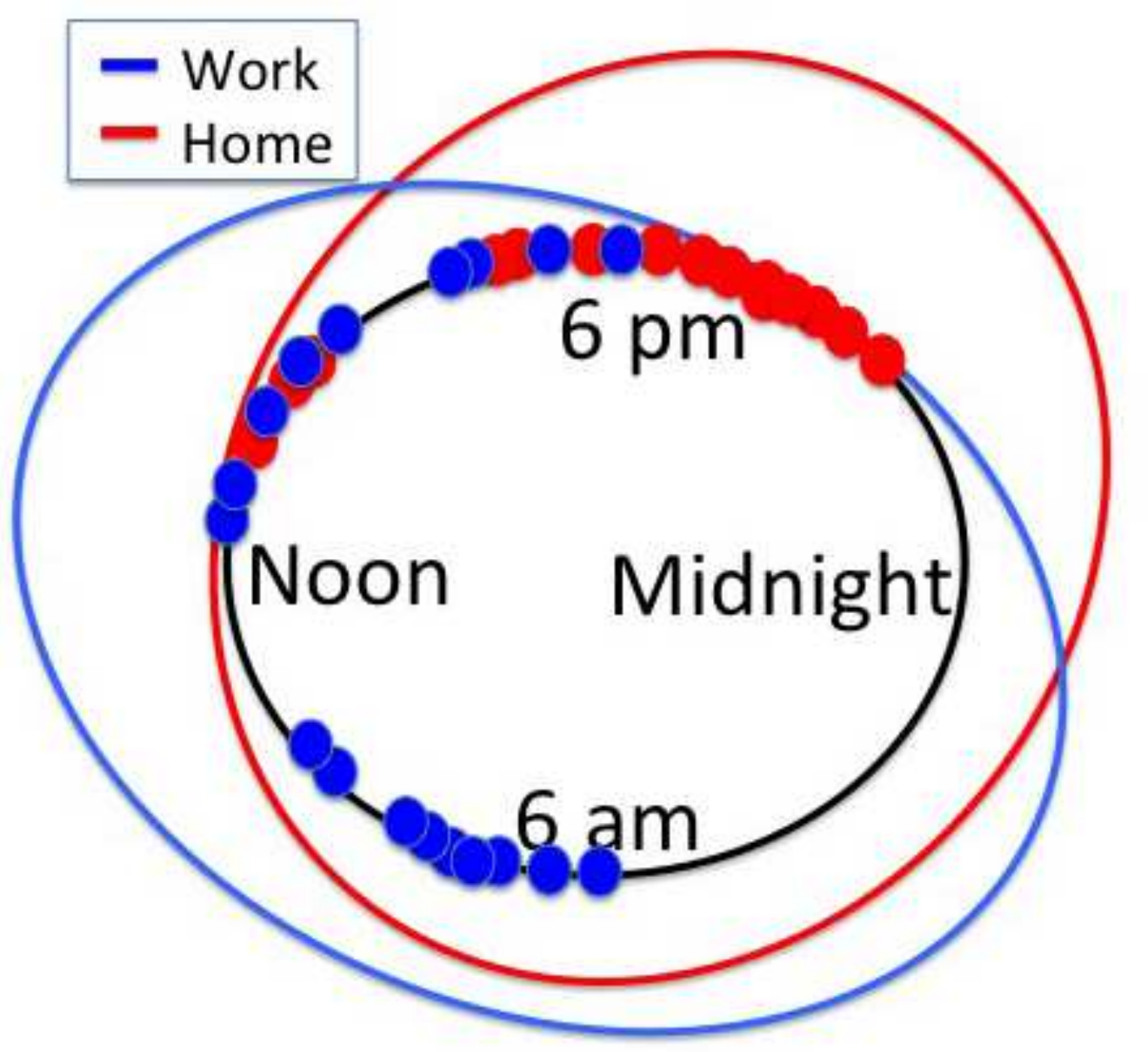}	
		\label{fig:UserTemporalBehavior}	 
		}
	\end{subfigure}
	\begin{subfigure}[Social Behavior]{
	\centering	
\includegraphics[height=3.5cm, width=4.2cm]{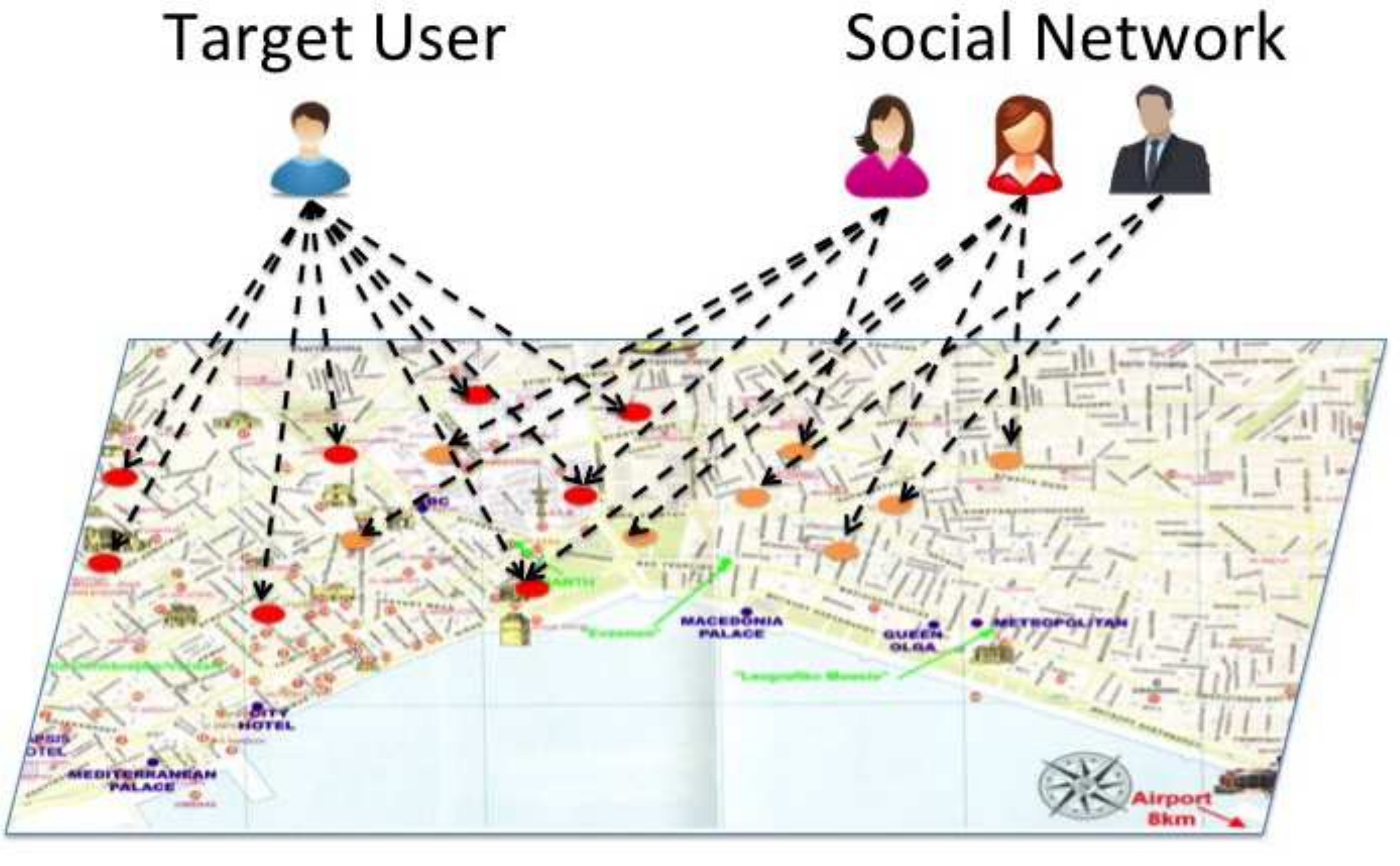}
		 \label{fig:UserSocialBehavior}	 
		 }
	\end{subfigure}			 
	\begin{subfigure}[Preference Dynamics]{
	\centering	
\includegraphics[height=3.5cm, width=4.2cm]{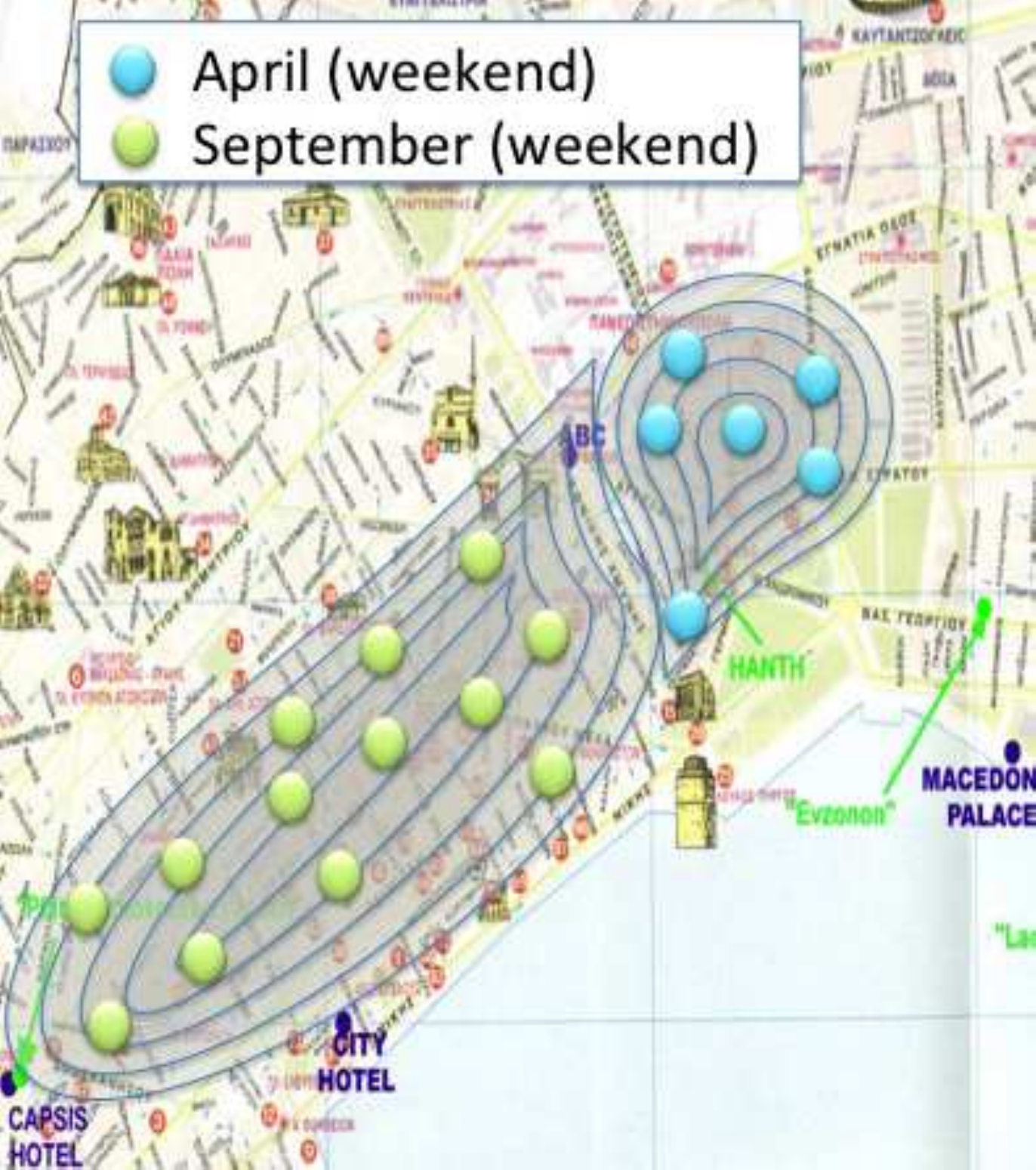}
		 \label{fig:PreferenceDynamics}	 
		 }		 
	\end{subfigure}		
\caption{(a) Spatial distribution of one user's check-ins close to home or work, (b) temporal distribution of check-ins where both lines indicate the proportional probability of being in one of these two states, (c) the social behavior indicate the influence of the social network to the target user, and (d) preference dynamics evolution between two months.} 
\label{fig:BehaviorDistributions} 
\end{figure*}
In this paper, we present a technique that considers social, geographical and temporal influence, along with users' preference dynamics, in a unified model, embedding eight relational graphs into a single latent space. 

\spara{Spatial-based Behavior}. Recent research points out that there is a spatial pattern on users' check-in behavior during daily activities~\cite{Cho2011}. In particular, users who check-in a location within a region have high probability to attend locations that are proximate. For example, a user located close to work or home has higher probability to visit a proximate location, rather than one in a long distance~\cite{Kefalas2017}. We claim what users tend to perform a sequence of activities during the same time period within a region which is related to a task. For example, users who want to do shopping usually check-in a mall, a supermarket, a grocery, or a  bakery located close to home, as shown in Figure~\ref{fig:UserSpatialBehavior}. Also, during weekends people perform multiple check-ins in clubs and restaurants that are close to each other. Thus, spatial proximity should be considered as a repeated geographical pattern. Also, the relation of those locations can be considered as a route of locations that are related to one activity. 

\spara{Temporal-based Behavior}. Usually, users maintain a fixed daily program in their activities and the check-ins they perform. Thus, on weekdays a user performs check-ins at locations close to work $6-5$, whereas from $5$ p.m. until midnight she check-ins at locations close to home, as shown in Figure~\ref{fig:UserTemporalBehavior}. This pattern is reversed in weekends because users tend to check-in bars and restaurants. Several works aim to model this behavior by focusing on temporal drifting but they ignore explicit and implicit contextual information~\cite{Lu2010a,Vasuki2010} related to users activities.

\spara{Preference Dynamics}. Users tend to change behavior during time which makes this problem even harder. For example, a user may attend clubs every weekend of September and cinemas or restaurants during April weekends, as shown in Figure~\ref{fig:PreferenceDynamics}. In both cases, the same user alternates his check-in behavior, which should be taken into account. This preference evolution according to~\cite{Xiong2010,Kefalas2017} may be due to: \begin{inparaenum}[1)] \item {\it New POIs exploration:} contrary to ordinary check-in behavior, users tend to visit new locations, \item {\it User experience:} users will choose a location according to locations in which he had a pleasant experience in the past, \item {\it Popularity:} some locations tend to be more popular during a time interval rather than the rest of the year, \item{\it Social influence:} friends' opinion bias users' decisions (as shown in Figure~\ref{fig:UserSocialBehavior}). Thus, users tend to examine their friends' evaluations over locations before attending and follow their lead. \end{inparaenum}

\spara{Motivation and Contribution}. Motivated by users' behavioral pattern in this section we summarize the limitations of existing approaches: 
\begin{itemize}
	\item many methods consider POIs as conventional nodes and do not capture the spatial dimension and the proximity among them,
	\item other works consider geographical influence but ignore preference dynamics tackling the over all accuracy,
	\item methods that capture temporal dynamics do not treat simultaneously the spatial dimension,
	\item finally, models that consider both spatial and temporal behavior ignore the preference evolution.
\end{itemize}	
Thus, there is a need to consider all aforementioned factors in a unified model which allows to further understand users' behavior and personalize the recommendations. The contributions of our work are as follows:
\begin{itemize}
	\item we present a probabilistic weighting strategy over the $8$ information graphs to overcome sparsity, 
	\item we propose two novel algorithms to extract the routes and the stay points out of the past history check-ins,
	\item we introduce a novel graph-based approach that jointly learns users' and POI embeddings from these weighted graphs into the same latent space and provides personalized POI recommendations,
	\item our approach extends the LINE model~\cite{Tang2015} and learn the embeddings of large unipartite and bipartite graphs simultaneously into a low dimensional space,
	\item we experimentally evaluated our model measuring the accuracy of POI recommendations for $i$) all the users, $ii$) the cold-start users, and $iii$) the cold-start locations. 
\end{itemize}	

\spara{Extensions beyond the Conference Version}. This work is an extended version of our work presented at the 5-$th$ IEEE International Conference on Data Science and Advanced Analytics (DSAA 2018)~\cite{DSAA2018ckpm}. This journal version contains several enhancements with respect to the conference paper. The most significant changes are summarized below:  
\begin{itemize}
\item we introduce two novel networks: $i$) \textit{Stay Points}, representing locations the user stayed the most, and $ii$) \textit{Routes}, the path followed when visiting POIs,
\item we incorporate two novel algorithms to build the aforementioned information networks which we weigh according to their importance,
\item we jointly capture users' and POIs sequential dynamics, 
\item the performance evaluation section has been extended significantly by studying two important topics in the domain related to the cold-start problem for both users and locations, and 
\item we compare our approach against additional state-of-the-art methods in terms of accuracy.
\end{itemize}

\spara{Roadmap}. The rest of the paper is organized as follows. Section~\ref{sec:Related work} summarizes the related work, whereas Section~\ref{sec:PPD} describes the preliminaries and the problem definition. In Section~\ref{sec:Proposed Approach}, we illustrate the model structural parts in details, while in Section~\ref{sec:Evaluation}, we present the findings of our experimentation. Finally, Section~\ref{sec:Conclusion} concludes this paper.

%% file: related.tex
\section{Related work}
\label{sec:Related work}

In this section, we discuss research conducted related to POI recommendation. In particular, we analyze how the prior work exploits $n$-dimensional networks to over come sparsity and cold start problems. These networks include social ties, geographical proximity, temporal distance, and preference dynamics of users' check-in past history.

\spara{POI Recommendations}.
The lack of direct relation between trajectory points and users' preferences derive from their check-in records tackled research to that direction. Recent years with the raise of LBSNs, users are able to check-in locations which has resulted to an anonymous access of their data for research purposes. To this context there has been a lot of attention in recommendation models that use such information. Most of related work uses Collaborative Filtering (CF)~\cite{Yuan2013}, Content-based Filtering (CB)~\cite{Meng2013,Gao2015} or hybrid~\cite{Kefalas2017,Xie2016,Yang2017} approaches to learn users' preferences over attended POIs and makes predictions for unvisited locations. The former approaches offer recommendations based on the assumption that users who visit the same POIs are most likely to visit the same locations in the future. For instance, Yuan et al. exploit the similarity among users through the check-in history and use collaborative filtering~\cite{Yuan2013}. On the other hand, content-based approaches use additional information related to users or POIs to tackle sparsity problem. Similarly, Gao et al.~\cite{Gao2015} exploit the content information of LBSNs by investigating the types of content information related to POI attributes along with check-in records. To overcome problems each method faces separately, such as: $i$) treating POIs as nodes and ignoring the geographical proximity, and $ii$) missing additional dimensions such as social influence or temporal evolution, hybrid approaches were introduced. Bellow we discuss in detail models that use additional information networks.

\spara{Social Influence}.
Since check-in records cannot always overcome the sparsity and cold-start problems, many approaches use social ties following the assumption that users tend to follow their friends' lead~\cite{Kefalas2017}. For example, Li et al.~\cite{Li2016} distinguish three types of friendships, that are $i$) linked, $ii$) co-located, and $iii$) proximate friends and exploit their check-in records through a unified framework. First, this model learns the common POIs that the target user and all three types of friends check-in the past. Then, it uses matrix factorization to minimize two loss functions over the learned POIs to personalize recommendations. Similarly, Zhang et al.~\cite{Zhang2014} introduced another unified model named LORE (Location Recommendations) that combines sequential patterns with social and geographical influence. This model exploits the sequential influence of POIs over users' records through a dynamic graph with Additive Markov Chain (AMC). Finally, it combines all aforementioned influences into one model with product fusion rule equation.

\spara{Geographical Influence}.
To further enhance the knowledge about users and eliminate sparsity, many methods use geographical information. Unfortunately, some of them treat POIs as conventional items~\cite{Liu2014} and miss the geographical proximity between locations. On the other hand, recent studies~\cite{Wang2016,Yang2017,Yin2017,Wang2017} consider the geographical proximity and treat them as `spatial items'. In particular, Wang et al.~\cite{Wang2016} focused on the importance of sequential influence of spatial items a user purchases, and proposed a \textit{Sequential PersOnalized item REcommender} system (SPORE) that fuses the sequential influence of spatial items and the preferences of a target user into the same latent space through a probabilistic topic-region unified model. Extending previous work related to spatial item recommendation, another probabilistic model~\cite{Wang2017} was introduced that jointly correlates the geographical influence, item attributes, and users' reviews into one unified framework named LSARS. Both models support the claim that users are willing to interact with proximate items, thus they use geographical influence jointly with additional factors.

Yin et al.~\cite{Yin2017} claimed that users keep the same preferences either being in their hometown or when they visit a new region. The authors used spatial attributes of POIs to alleviate sparsity and cold-start problems for out-of-town users. The geographical influence, such as region attributes, is used as auxiliary information to auto-correlate users with new locations. That model, named SH-CDL, jointly learns these attributes along with additive representations of users' check-in preferences. This way they search for proximate items close to users' past preferences in other regions. In the same content, Yang et al~\cite{Yang2017} developed a \textit{Preference And Context Embedding} (PACE) model that jointly learns the embeddings of users and locations. This hybrid model bridges Semi-Supervised Learning (SSL) with Collaborative Filtering (CF) in a unified way.
 
\spara{Temporal Influence}.
Recent state-of-the-art studies indicate that there is a periodicity on users' behavior which should also be encountered along with all aforementioned factors~\cite{Kefalas2017,Zhao2017,Liu2017}. Yuan et al.~\cite{Yuan2013} examined the spatiotemporal behavior during different hours of day claiming that users tend to visit specific locations at different time periods. They proposed a unified model that combines CF and Bayes rule to provide POI recommendations considering both proximity and periodicity. Similarly, Kefalas et al.~\cite{Kefalas2017} proposed another hybrid model that combines the CF and CB in a unified way exploring i) the proximate users' preferences, ii) the textual influence alternation within time periods, and iii) the preference dynamics evolution. Results indicate an evolution of users' check-in behavior, since they are highly influenced by all factors combined. The results pointed an incremental robustness of the precision against models considering each factor separately. In the same direction, Liu et al.~\cite{Liu2017} developed another hybrid spatiotemporal-aware model that learns the jointly representations of users, spatiotemporal patterns and POIs. 

On the other hand, following the embedding models like word2vec~\cite{Mikolov2013}, Zhao et al.~\cite{Zhao2017} presented Geo-Temporal sequential embedding rank (Geo-Teaser) that combines both $i$) the temporal POI embeddings, and $ii$) the spatial hierarchical pairwise ranking. In particular, the first half model learns POI representations based on a temporal POI embedding model which uses users' daily check-in history as a sequence. The second half model ranks POIs based on geographical information in a hierarchical pair-wise preference. Similarly, Xie et al.~\cite{Xie2016} presented a unified graph-based model named GE that jointly learns POI embeddings by capturing the sequential effect, the spatial influence, the periodicity and the semantics into the same latent space.

In contrast to existing work, we propose a model that considers all aforementioned factors into one unified model. In particular, we apply a probabilistic strategy to weigh the importance of edges over eight information graphs. Then, we introduce a graph-based approach that jointly learns users' and POIs embeddings from these weighted graphs into the same latent space and provides personalized POI recommendations. Furthermore, we examine the performance of our approach against \textit{cold-start users} and \textit{cold-start POIs} problems in terms of accuracy. The proposed technique shows significant performance improvement in comparison to existing state-of-the-art methods. To the best of our knowledge, this is the first attempt in literature to face both problems simultaneously.

%% file: preliminaries.tex
\section{Preliminaries and Problem Definition}
\label{sec:PPD}

\begin{figure}[!h]
	\centering			
\includegraphics[height=6cm, width=9cm]{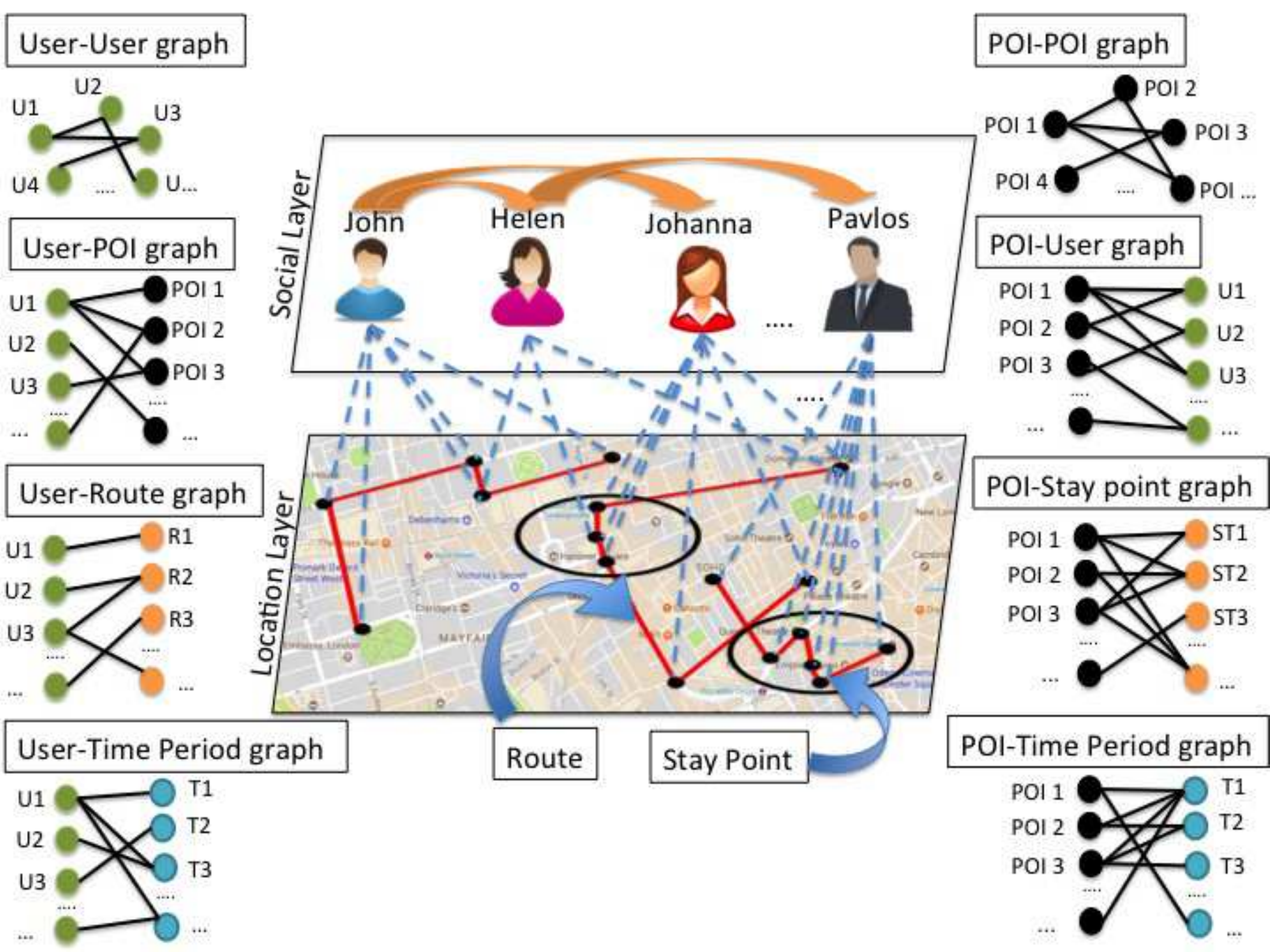}
\caption{An example of participating graphs in an LBSN.} 
\label{fig:ToyExample} 
\end{figure}

In this section, we present the problem definition in detail and also discuss some fundamental concepts related to our research. Figure~\ref{fig:ToyExample} depicts all participating networks, whereas Table~\ref{tab:Symbols and Notations} presents the most frequently used symbols in the sequel. 
Some fundamental definitions follow.

\begin{mydef}{\bf (POI):}\label{def:POI} is a unique location in which users checked-in. A POI is represented as a tuple: $<$$l_{id}$, longitude, latitude$>$.
\end{mydef}

\begin{mydef}{\bf (Check-in):}\label{def:Check-in} is a self-report positioning of a user $u$, in a location $l$, at time $t$, represented as a tuple: $c_i$=$<$u,l,t$>$. One check-in can be performed only by one user, but the same user may have multiple check-ins recorded in the profile $c_{u_i}=\{<l_i,t_i>,...,<l_j,t_j>\}$.
\end{mydef}

\begin{mydef}{\bf (Time Period):}\label{def:TimePeriod} is defined as distinct time intervals of the entire dataset, divided in equal size bins of days, weeks or months. Each period contains the check-ins of all users performed during that time interval $\Delta T$.
\end{mydef}

\input{LaTeX/Notations.tex}

\begin{mydef}{\bf (User-User graph):}\label{def:User-User graph} is a unipartite graph that describes the social network of the users. This graph is denoted as $\mathcal{G}_{uu}=(U \cup V,\mathcal{W}_{uv})$, where $U$ and $V$ are sets of users, and $\mathcal{W}_{uv}$ is the set of the weighted edges among them. This network is an undirected graph of friendship relations among users. Thus, the bidirectional connection between the vertices is described as $<u,v>$ = $<v,u>$ and the weight is defined as the fraction of 1 over the number of users ($v$), who are friends with user $u$:
\begin{equation}
	\begin{aligned}
		\mathcal{W}_{uv} = \frac{1}{\sum\limits_{i=1}^{n}{|v_i|}}
		\label{eq:User-User-weight}
	\end{aligned}
	\notag
\end{equation}
\end{mydef}

\begin{mydef}{\bf (User-POI graph):}\label{def:User-POI graph} is a directed bipartite graph which indicates relation between users and locations for the entire check-in history. In particular, it defines the importance of one location for a target user against all the other locations. This graph is denoted as $\mathcal{G}$$_{ul}=(U \cup L,$ $\mathcal{W}$$_{ul})$, where $U$ is the set of users, $L$ is the set of locations, and $\mathcal{W}_{ul}$ is the set of the weighted edges among them. The weights are computed as the fraction of the number of times a user $u$ visited location $l$ against all over all her check-ins:

\begin{equation}
	\begin{aligned}
		\mathcal{W}_{ul} = \frac{\sum\limits_{\forall c_{u_i} \in l_i}{c_{u_{i,l}}}}{\sum\limits_{\forall c_{u_i} \in L}{c_{u_{i,L}}}}
		\label{eq:User-POI-Weight}
	\end{aligned}
	\notag
\end{equation}
\end{mydef}

\begin{mydef}{\bf (User-Time period graph):}\label{def:User-TimePeriod graph} is a directed bipartite graph representing the interaction of a user at a time period. This graph is denoted as $\mathcal{G}$$_{ut}=(U \cup T,$ $\mathcal{W}$$_{ut})$, where U is the set of users, T is the set of time periods, and $\mathcal{W}$$_{ut}$ is the set of the weighted edges among them. The weights are computed as the fraction of the number of each user check-ins during a time period against all the check-ins s/he has made:

\begin{equation}
	\begin{aligned}
		\mathcal{W}_{ut} =\frac{\sum\limits_{\forall c_{u_i} \in t_i}{c_{u_{i,t}}}}{\sum\limits_{\forall c_{u_i} \in T}{c_{u_{i,T}}}} 
		\label{eq:User-Time-Weight}
	\end{aligned}
	\notag
\end{equation}
\end{mydef}

\begin{mydef}{\bf (Route):}\label{def:Route} is the sequence of POIs each user attended during a time period $\Delta T$. For example $\{t_1 : l_1 \to l_2 \to l_3 \}$ indicates that user u moved from location $l_1$ to location $l_2$ and then ended his route at location $l_3$ during time period $t_1$. To extract the routes of each user, we first short the check-ins according to time. Then, we split the dataset into time periods and for each user we construct the route sequence during that time interval. 
\end{mydef}

\begin{mydef}{\bf (User-Route graph):}\label{def:User-Route graph} is a directed bipartite graph which emphasizes the importance of a route for each user. First, we extract the routes from given POIs (as described in Definition~\ref{def:POI}) with Algorithm~\ref{algo:RouteExtraction}. 

\input{LaTeX/RouteExtractionAlgorithm.tex}

Then, we correlate each user from $U$ to one route from the set $R$. This relation is represented as $\mathcal{G}$$_{ur}=(U \cup R,$ $\mathcal{W}$$_{ur})$. The weight of the edge between two nodes is computed as the number of times one user followed one particular route to the total number or routes made by the same user during all time periods:

\begin{equation}
	\begin{aligned}
		\mathcal{W}_{ur} = \frac{\sum\limits_{\forall l_{u_{i,r}} \in t_i}{c_{u_{i,t}}}}{\sum\limits_{\forall l_{u_{i,r}} \in T}{R_{u_{i,T}}}}
		\label{eq:User-Route-Weight}
	\end{aligned}
	\notag
\end{equation}
\end{mydef}

\begin{mydef}{\bf (POI-POI graph):}\label{def:POI-POI graph} is a bidirectional bipartite graph pointing the spatial proximity between two locations. In particular, two locations $l$ and $l'$ are connected with an edge, if and only if one user $u$ check-ins both of them during a time period $t$ and a distance range $Rg$. According to this assumption we build our graph denoted as $\mathcal{G}$$_{ll'}=(L \cup L,$ $\mathcal{W}$$_{ll'})$, where $L$ is the set of locations, and $\mathcal{W}_{ll'}$ is the set of the weighted edges between two location nodes considering their geographical proximity. This weight is computed as:
\begin{equation}
	\begin{aligned}
		\mathcal{W}_{ll'} = 1 - \frac{(geodist_{l,l'})}{Rg} 
		\label{eq:User-POI-POI}
	\end{aligned}
	\notag
\end{equation}
where $geodist_{l,l'}$ computes the geographical distance between two locations.
\end{mydef}

\begin{mydef}{\bf (POI-User graph):}\label{def:POI-User graph} is also a directed bipartite graph which points the relation between a location and a user. The main difference in that graph is the weighting strategy we follow, since the influence of a location to a user differs from the previous approach. In particular, this graph is denoted as $\mathcal{G}$$_{lu}=(U \cup L,$ $\mathcal{W}$$_{lu})$, where L is the set of locations, and U is the set of users, and $\mathcal{W}$$_{lu}$ is the set of the weighted edges among them. The weight is computed as the fraction of the number of times a location $l$ is visited by a user $u$ to the number of all users checked-in that location:

\begin{equation}
	\begin{aligned}
		\mathcal{W}_{lu} = \frac{\sum\limits_{\forall c_u \in l_i}{c_{u}}}{\sum\limits_{\forall c_U \in l_i}{c_{U}}}
		\label{eq:POI-User-Weight}
	\end{aligned}
	\notag
\end{equation}
\end{mydef}

\begin{mydef}{\bf (POI-Time Period graph):}\label{def:POI-Timestamp graph} is a directed bipartite graph which indicates the importance of a location during a time period. This graph is denoted as $\mathcal{G}$$_{lt}=(L \cup T,$ $\mathcal{W}$$_{lt})$, where L is the set of locations, T is the set of Time periods, and $\mathcal{W}$$_{lt}$ is the set of the weighted edges among them. The weight is computed as the fraction of the number of the check-ins performed at a location $l$ by all users, during a time period $t_i$, against the total number of check-ins during all time periods at the same location:

\begin{equation}
	\begin{aligned}
		\mathcal{W}_{lt} = \frac{\sum\limits_{\forall c_U \in t_i}{|n_{lt}|}}{\sum\limits_{\forall c_U \in T}{|n_{lt}|}}
		\label{eq:POI-Time-Weight}
	\end{aligned}
	\notag
\end{equation}
\end{mydef}

\begin{mydef}{\bf (POI-Stay Points graph):}\label{def:POI-StayPoints graph} is a bidirectional bipartite graph which describes the significance of some locations for a user. In daily schedule users follow some routes and spend some time in each location of that route sequence. The elapsed time of each location indicates the importance of this location to this particular user. Thus, the more time is spent on a location, the higher the importance of that location is. To extract these important locations, denoted as `stay points' ($st_i$), we use Algorithm~\ref{algo:ST}. 

\input{LaTeX/StayPointsAlgorithm.tex}

Then we construct the $\mathcal{G}$$_{lst}=(L \cup ST,$ $\mathcal{W}$$_{lst})$ graph, where L is the set of locations, ST is the set of stay points, and $\mathcal{W}$$_{lst}$ is the set of the weighted edges among them. The weight is computed as the fraction of the number of times one location is considered as a stay point in a route, against the total number of times this location is considered as stay point in all routes:

\begin{equation}
	\begin{aligned}
		\mathcal{W}_{lst} = \frac{\sum\limits_{\forall st_i \in r}{l}}{\sum\limits_{\forall st_i \in R}{l}}
		\label{eq:POI-StayPoint-Weight}
	\end{aligned}
	\notag
\end{equation}
\end{mydef}

\noindent
{\bf Problem Definition:} {\it ``Given a user $u$, a location $l$ and a time instance $t$ (expressed as $Q(u,l,t)$), and the check-in history, predict the top-$N$ unvisited proximate POIs to that target user.''}
\label{ProblemDefinition}

%% file: LaTeX/Notations.tex
\begin{table}[!b]
\caption{Frequently used symbols.}
\label{tab:Symbols and Notations}
\centering
\begin{tabular}{| l | l |} \hline\hline
Symbol & Description \\ \hline \hline
$U,L,T,R,C$	    & set of users $U=\{u_1,...,u_n\}$, 				    \\		
			    & set of locations $L=\{l_1,...,l_m\}$, 			    \\ 
			    & set of Time intervals $T=\{t_1,...,t_k\}$, 		    \\
			    & set of Routes $R=\{r_1,...,r_p\}$, 				    \\	
			    & set of Check-ins $C=\{c_1,...,c_o\}$ 				    \\ \hline
$c_{u_i}$		& user's check-in								        \\	\hline
$l_{u_{i,r}}$	& location belong to user's route				        \\ \hline
$r_{num}$		& number of POIs in routes						        \\ \hline
$\Delta T$		& time interval									        \\	\hline
$S$, $N$		& number of samples, number of negative samples		    \\	\hline
$e_{i,j}$	    & edge between two nodes 							    \\	\hline
$E$			    & set of edges $e_{i,j}$ over each graph 			    \\	\hline
$w_{i,j}$	    & weighted edge $e_{i,j}$ 							    \\	\hline
$W$		    	& set of weights $w_{i,j}$ over each graph 			    \\	\hline
$\mathcal{S}$   & an edge $e_{i,j}$	sampled from $W$					\\	\hline
\end{tabular}
\end{table}

%% file: LaTeX/RouteExtractionAlgorithm.tex
\begin{algorithm}[!t]
\DontPrintSemicolon
\KwIn{$C$: Set of users' check-ins history, \\
\qquad\quad	$\Delta T$: time interval \\ }
\KwOut{$R$: a list of routes of all users'}

Sort C based on timestamp\\
\label{algorithmline1.1}
Create T periods with time interval $\Delta T$\\
\label{algorithmline1.2}
Create list of Routes R\\
\label{algorithmline1.3}

\For(\tcp*[f]{\small{For every route $t_i$ $\in$ $T$}}){(i=0, i==T, i++)}{
	n $\gets$ 0;\;						\label{algorithmline1.4}
    \While(\tcp*[f]{\small{While $c[n]$ $\in$ $T_i$}}){($T_{start} \leq c_{u_i}[n] \le T_{end}$)}{
  	T[i] $\gets$ $c_{u_i}[n]$;\; 				\label{algorithmline1.5}
	n ++;\;							\label{algorithmline1.6}
	}
	q $\gets$ 0;\;						\label{algorithmline1.7}
	\For(\tcp*[f]{\small{For every route $l$ $\in$ $T_i$}}){(j=0, j==sizeof.T[i], j++)}{
		\If(\tcp*[f]{\small{By the same user $u$}}){($u_i[j+1] == u_i[j]$)}{
			R[q].add($c_{u_i}[n]$)		\label{algorithmline1.8}	
		}	
		\Else{		
			q ++;\;					\label{algorithmline1.9}
			}		
	}			
    }
    
\Return R\;							\label{algorithmline1.10}

\caption{{\sc Route Extraction from Check-ins for each time period}}
\label{algo:RouteExtraction}
\end{algorithm}

%% file: LaTeX/StayPointsAlgorithm.tex
\begin{algorithm}[!t]
\DontPrintSemicolon
\KwIn{$R$: Set of users' routes }
\KwOut{$SP$: a list of stay points of each route}
 
\For(\tcp*[f]{\small{For every route $r_i$ $\in$ $R$}}){(i=0; i $<$ p; i++;)}{	
 	\For(\tcp*[f]{\small{For every location $l_j$ $\in$ $r_i$}}){(j=1; j $<$ $r[i]_{num}-1$; i++;)}{	
  		Duration[j] $\gets$ $l_j.t_i - l_{j-1}.t_i$;\;	\label{algorithmline2.1}
		}
		
   	StayPoint $\gets$ $l$.max(Duration); \tcp*[r]{\small{Find $l_j$ with maximum elapsed time in $r_i$}}						\label{algorithmline2.2}
 	ST[i].add(StayPoint);\;					\label{algorithmline2.3}
}
    
\Return ST\;								\label{algorithmline2.4}
\label{algorithmline17}

\caption{{\sc Stay Points Extraction from users' routes}}
\label{algo:ST}
\end{algorithm}

%% file: proposed.tex
\section{Proposed Approach}
\label{sec:Proposed Approach}
Next, we present RELINE, an optimized solution for jointly learning the graph embeddings of different information networks in the same latent space and we propose a unified framework for POI recommendations.

\subsection{Learning Embeddings in a Bipartite Graph}
\label{sec:Learn embeddings in a bipartite graph}

Nodes that are directly connected to an edge $e_{i,j}$ and weight $w_{i,j}$, consist the \textbf{\textit first-order proximity}, that is the local pairwise closeness between two nodes. On the other hand, nodes that share many connections but they are not directly related to an edge, they belong to the same neighborhood, thus, there are most likely to be similar to each other. These nodes consist the \textbf{\textit second-order proximity}, that is the similarity between two unlinked nodes according to their network structure. To extract this kind of proximity on uni-partite graphs, the LINE model~\cite{Tang2015} learns the embeddings of large graphs into a low dimensional space. With this work, we extend this model to learn the embeddings over bipartite graph nodes. Moreover, our model can be applied into all kind of bipartite graphs i.e., directed/undirected, weighted/unweighted, or combinations of them.

Given two disjoinτ sets $\mathcal{G} =(\mathcal{S}_{A} \cup \mathcal{S}_{B},\mathcal{W})$, the nodes in $\mathcal{S}_{A}$ that share many connections with $\mathcal{S}_{B}$ but are not directly connected to each other, are most likely to have the same distributions. The conditional probability of one node $v_j \in \mathcal{S}_{B}$ is generated through node $v_i \in  \mathcal{S}_{A}$ such as:

\begin{equation}
	\begin{aligned}
		p(v_j | v_i) = \frac
			{exp(\vv{v}_j^T \cdot \vv{v}_i)}
			{\sum\limits_{u_{k} \in \mathcal{S}_{B}} exp(\vv{v}_k^T \cdot \vv{v}_i)}
		\label{eq:Conditional probability}
	\end{aligned}
\end{equation}

\noindent where the embeddings vectors of vertices $v_i$, and $v_j$ are represented as $\vv{v}_i$ and $\vv{v}_j$, respectively. Thus, for each node $v_i \in \mathcal{S}_{A}$, Equation~(\ref{eq:Conditional probability}) defines the conditional distribution $p(\cdot | u_i)$ to all the corresponding nodes in the set $\mathcal{S}_{B}$. Then, for each edge there is a weight which implies the strength of this tie. To retain the proximity of the unlinked nodes in $\mathcal{S}_{A}$, we let the described conditional distribution to approximate the empirical distribution $\hat{p}(\cdot|v_i) = \frac{w_{i,j}}{\sum{w_{i,k}}}$ with the following objective function:
\begin{equation}
	\begin{aligned}
O = \sum\limits_{u_{i}\in \mathcal{S}_{A}}\lambda_i d(\hat{p}(\cdot|v_i), p(\cdot|v_i))
		\label{eq:BOF}
	\end{aligned}
\end{equation}
\noindent where $d(\cdot | \cdot)$ denotes the Kullback-Leibler divergence of the conditional and the empirical distributions, and $\lambda_i$ is a regularization parameter to tune the significance of node $v_i$. For simplicity, we set this parameter equal to the out-degree of each node. Thus, Equation~(\ref{eq:BOF}) corresponds to the minimization of the following objective function:

\begin{equation}
	\begin{aligned}
	O = - \sum\limits_{e_{i,j} \in W} w_{i,j} \text{ log } p(v_j|v_i)
		\label{eq:OBOF}
	\end{aligned}
\end{equation}

\noindent The vectors $\{\vv{v}_i\}_{i=1..\mathcal{S}_{A}}$ and $\{\vv{v}_j\}_{j=1..\mathcal{S}_{B}}$ that minimize Equation~(\ref{eq:OBOF}) are the low-rank nodes representations in $\mathbb{R}^d$. 

\subsection{Model Optimization trough Negative Sampling}
\label{sec:Learn embeddings in a Bipartite graph}

To avoid the calculation of the conditional probability $p(\cdot|v_i)$ which needs the summation of the entire set of nodes in Equation~(\ref{eq:OBOF}), we apply negative sampling (NEG)~\cite{Mikolov2013} over each edge. In particular, we use the noisy distribution of each edge individually to sample $N$ negative edges as described in the following objective function:

\begin{equation}
	\begin{aligned}
\text{ log } \sigma(\vv{v}_j^T \cdot \vv{v}_i) + \sum\limits_{h=1}^{N} W_{u_n \sim P_{n}(u)} [\text{ log } \sigma(-\vv{v}_n^T \cdot \vv{v}_i)]
		\label{eq:FOBOF}
	\end{aligned}
\end{equation}

\noindent where $\sigma(x) = 1/1+exp(-x)$ is the sigmoid function with output values [0-1], and $P_n(v \propto d_{u}^{3/4})$ is the noise distribution in which the negative samples are chosen with a unigram distribution empirically tuned in~\cite{Mikolov2013}, such that each node occurrence in the set is independent of all other node occurrences. Thus, selecting a node as a negative sample is related to the out-degree in that power. To further improve the solution of Equation~(\ref{eq:FOBOF}), we apply `Hogwild' algorithm~\cite{Niu2011} based on the asynchronous stochastic gradient descent (ASGD). In particular, each time an edge $e_{i,j}$ is sampled, we calculate the gradient of node $v_i$ over the corresponding embedding vector $\vv{v}_i$ as follows:

\begin{equation}
	\begin{aligned}
	\frac{\partial O}{\partial \vv{v}_i}  = 
	w_{i,j} \cdot \frac{\partial \text{ log }p(v_j|v_i)}
				       {\partial \vv{v}_i}
		\label{eq:Gradients}
		\notag
	\end{aligned}
\end{equation}

Notice that the gradient of node $v_i$ is multiplied with the weight related to that edge. Thus, tuning the learning rate of the model, may cause problems due the valiance of the weights. On the one hand, {\bf`overfitting'} may occur to the gradients with large weights, if large learning rate is chosen according to edges multiplied with small weights values. On the other hand, {\bf`underfitting'} may occur to the gradients with small weights, if small learning rate is chosen for edges multiplied with large weights values. 

To balance the learning rate or our model, we adopt the sampling method presented in~\cite{Tang2015}. In particular, we sample a random edge $e_{i,j} \in [0-W_{sum}]$, where $W_{sum}$ denotes the sum of all weights in the particular network, and then we examine the interval in which the particular sampled edge falls into, i.e.$[\sum_{j=0}^{i-1} w_j, \sum_{j=0}^{i} w_j ]$. Finally, we draw a sampled edge using alias table according to~\cite{Li2014} which eventually reduces the complexity to $\mathcal{O}(1)$. Table~\ref{tab:Complexity} presents the over all complexity of edge sampling optimization. 

\begin{table}[ht]
	\centering
		\caption{Complexity analysis.}\label{tab:Complexity}
	\begin{tabular}[t]{|c|c|}
		\hline
		Sample edge from alias table		&$\mathcal{O}(1)$\\\hline	
		Negative sampling optimization		&$\mathcal{O}(N+1)$\\\hline
		Overall complexity				&$\mathcal{O}(N\cdot E)$\\\hline
	\end{tabular}
\end{table}

\vspace{-0.5cm}

\subsection{Joint Learning of Graph Dynamics}
\label{sec:Joint Embedding Learning of Graph Dynamics}
Given the input bipartite graphs, the next step is to integrate them into our model. Graphs that correspond to users' relations are: User-User, User-POI, User-Route, and User-Time Period. On the other hand, graphs that correspond to the location ties with other networks are: POI-POI, POI-User, POI-Important locations, and POI-Time period. We collectively integrate the embeddings of these graphs corresponding to users' and POIs ties, by minimizing the objective function:

\begin{equation}
	\begin{aligned}
	\small{O = \underbrace{O_{uu} + O_{ur} + O_{ut} + O_{ul}}_\text{user networks} + \underbrace{O_{ll} + O_{lst} + O_{lt} + O_{lu}}_\text{POI networks}}
		\label{eq:ObjectiveFunction}
	\end{aligned}
\end{equation}

\noindent where each of the above objective functions is computed as:

\scalebox{0.9}{
\hspace{-1.16cm}
\renewcommand{\arraystretch}{2.0}
    \setlength{\tabcolsep}{-0.09em}
	\begin{tabular}{ c c }
    $ O_{uu} = - \sum\limits_{e_{i,j} \in W_{uu}} w_{i,j} \text{ log } p(u_i|v_j)$,	\label{eq:Ouu} 
    &$ O_{ll} = - \sum\limits_{e_{i,j} \in W_{ll}} w_{i,j} \text{ log } p(l_i|l_j)$	\label{eq:Oll}      \\
    $ O_{ur} = - \sum\limits_{e_{i,j} \in W_{ur}} w_{i,j} \text{ log } p(u_i|r_j) $, \label{eq:Our}
    &$ O_{lst} = - \sum\limits_{e_{i,j} \in W_{lst}} w_{i,j} \text{ log } p(l_i|st_j)$	\label{eq:Olst} \\
    $ O_{ut} = - \sum\limits_{e_{i,j} \in W_{ut}} w_{i,j} \text{ log } p(u_i|t_j) $,	\label{eq:Out}
    &$ O_{lt} = - \sum\limits_{e_{i,j} \in W_{lt}} w_{i,j} \text{ log } p(l_i|t_j)$	\label{eq:Olt}      \\
    $ O_{ul} = - \sum\limits_{e_{i,j} \in W_{ul}} w_{i,j} \text{ log } p(u_i|l_j)$,	\label{eq:Oul}
    &$ O_{lu} = - \sum\limits_{e_{i,j} \in W_{lu}} w_{i,j} \text{ log } p(l_i|u_j)$	\label{eq:Olu}      \\
\end{tabular}
}

To minimize the objective function presented in Equation~(\ref{eq:ObjectiveFunction}), first we merge together edges of all unipartite and bipartite graphs, and then, in each step we update the model by sampling a new edge. The probability of sampling an edge corresponds to the weight related to that edge. This way our model walks through the heterogeneous bipartite graphs with respect to the inner and the outer vertices of the graphs and the weight influence. The training of the model is done jointly and dynamically as shown in Algorithm~\ref{algo:DynamicJoinTrain}.

\input{LaTeX/DynamicJointTraining.tex}

\subsection{Unified Model for POI Recommendations}
\label{sec:JDGE}

By the time all embeddings, presented in previous section, have been learned, and given a prediction request $Q(u,l,t)$, concerning a user $u$ in a location $l$ at a timestamp $t$, we project these values to the corresponding time period $t$, route $r$, and stay point $st$, with geographical range distance less than 10 Km. We claim that one user is willing to attend proximate locations. Thus, given a recommendation further than this range, the probability of attending is very small. Then, we rank a list with the top@$n$ unvisited candidate POIs for that user in that distance. The prediction score for each of the unvisited location is computed as:

\begin{equation}
\noindent\resizebox{0.47\textwidth}{!}{$
Q(u,l,t) =  	\underbrace{\alpha \cdot (\vv{\textbf{u}}^T   \cdot 	\vv{\textbf{l}})}_\text{\tiny Social inf.} + 
				\underbrace{\beta \cdot (\vv{\textbf{r}}^T 	\cdot 	\vv{\textbf{l}})}_\text{\tiny Geographical inf.} + 
				\underbrace{\gamma \cdot (\vv{\textbf{t}}^T 	\cdot 	\vv{\textbf{l}})}_\text{\tiny Temporal inf.} + 
				\underbrace{\delta \cdot (\vv{\textbf{st}}^T  \cdot 	\vv{\textbf{l}})}_\text{\tiny Preference dynamics}	$}
		\label{eq:RELINE}
\end{equation}

\noindent where $\vv{\textbf{u}}$ is the embedding of user $u$, $\vv{\textbf{l}}$ is the embedding of location $l$, $\vv{\textbf{r}}$ is the embedding of the route $r$ this check-in belongs to, $\vv{\textbf{t}}$ is the embeddings of the time period $t$ in which the particular check-in was made, and $\vv{\textbf{st}}$ is the embedding of the stay points. Moreover, RELINE learns jointly the embeddings from different information networks in the same latent space. Thus, the learned POI embeddings $\vv{\textbf{l}}$ capture the information of all participated networks presented in the previous sections, such as $uu$, $ul$, $ll$ etc. This way, we aim to eliminate sparsity, simply by using additional information networks. In particular, we jointly learn the dynamics of the {\it social influence} $(\vv{\textbf{u}}^T \cdot \vv{\textbf{l}})$, the {\it the geographical influence} $(\vv{\textbf{r}}^T \cdot \vv{\textbf{l}})$, the {\it the temporal influence} $(\vv{\textbf{t}}^T \cdot \vv{\textbf{l}})$, and the {\it user's preference dynamics} $(\vv{\textbf{st}}^T \cdot \vv{\textbf{l}})$, simultaneously. Finally, $\alpha$, $\beta$, $\gamma$, and $\delta$ are regularization parameters that define the importance of each information network separately into our model.

%% file: LaTeX/DynamicJointTraining.tex
\begin{algorithm}[!ht]
\DontPrintSemicolon
\KwIn{$\mathcal{G}_{uu}$,$\mathcal{G}_{ul}$, $\mathcal{G}_{ur}$,$\mathcal{G}_{ut}$, $\mathcal{G}_{ll}$, $\mathcal{G}_{lu}$, $\mathcal{G}_{lr}$, $\mathcal{G}_{lt}$, S, N}
\KwOut{$\vv{\textbf{u}}$: User embeddings, $\vv{\textbf{l}}$: POI embeddings, \\
		\qquad\qquad $\vv{\textbf{r}}$: Route embeddings, $\vv{\textbf{t}}$: Time period embeddings\\
		\qquad\qquad $\vv{\textbf{st}}$: Stay point embeddings}

Initialize $\vv{\textbf{u}}$, $\vv{\textbf{l}}$, $\vv{\textbf{r}}$, $\vv{\textbf{t}}$, $\vv{\textbf{st}}$;\;\label{algorithmline3.1} 

\While(\tcp*[f]{Until the desired number of samples reached}){(iteration $\leq$ S)}{
At each iteration Draw N negative edges of the corresponding graph \label{algorithmline3.2} \;
$\mathcal{S}$($w_{uu})\in \mathcal{W}_{uu}$, and update $\vv{\textbf{u}}$; \label{algorithmline3.3} \;
$\mathcal{S}$($w_{ul})\in \mathcal{W}_{ul}$, and update $\vv{\textbf{u}}$ and $\vv{\textbf{l}}$;\label{algorithmline3.4} \;
$\mathcal{S}$($w_{ll})\in \mathcal{W}_{ll}$, and update $\vv{\textbf{l}}$;\label{algorithmline3.5} \;
$\mathcal{S}$($w_{lu})\in \mathcal{W}_{lu}$, and update $\vv{\textbf{l}}$ and $\vv{\textbf{u}}$;\label{algorithmline3.6} \;
$\mathcal{S}$($w_{ut})\in \mathcal{W}_{ut}$, and update $\vv{\textbf{u}}$ and $\vv{\textbf{t}}$;\label{algorithmline3.7} \;
$\mathcal{S}$($w_{lt})\in \mathcal{W}_{lt}$, and update $\vv{\textbf{l}}$ and $\vv{\textbf{t}}$;\label{algorithmline3.8}  \;
$\mathcal{S}$($w_{ur})\in \mathcal{W}_{ur}$, and update $\vv{\textbf{u}}$ and $\vv{\textbf{r}}$; \label{algorithmline3.9} \;
$\mathcal{S}$($w_{lst})\in \mathcal{W}_{lst}$, and update $\vv{\textbf{l}}$ and $\vv{\textbf{st}}$;\label{algorithmline3.10} \;
}		    
\Return $\vv{\textbf{u}}$,$\vv{\textbf{l}}$, $\vv{\textbf{r}}$,$\vv{\textbf{t}}$,$\vv{\textbf{st}}$ \label{algorithmline1.11}\;

\caption{\sc Dynamic Join Training of the embeddings}
\label{algo:DynamicJoinTrain}
\end{algorithm}

%% file: evaluation.tex
\section{Performance Evaluation}
\label{sec:Evaluation}
In this section, we focus on the performance of the proposed methodology. In particular, we compare our technique  against previous ones. The source code of our approach is available at \url{https://github.com/thedx4/RELINE}.

\begin{figure*}[!ht]
\begin{center}
\begin{minipage}{2.9cm}
\centerline{\includegraphics[width=3.1cm]{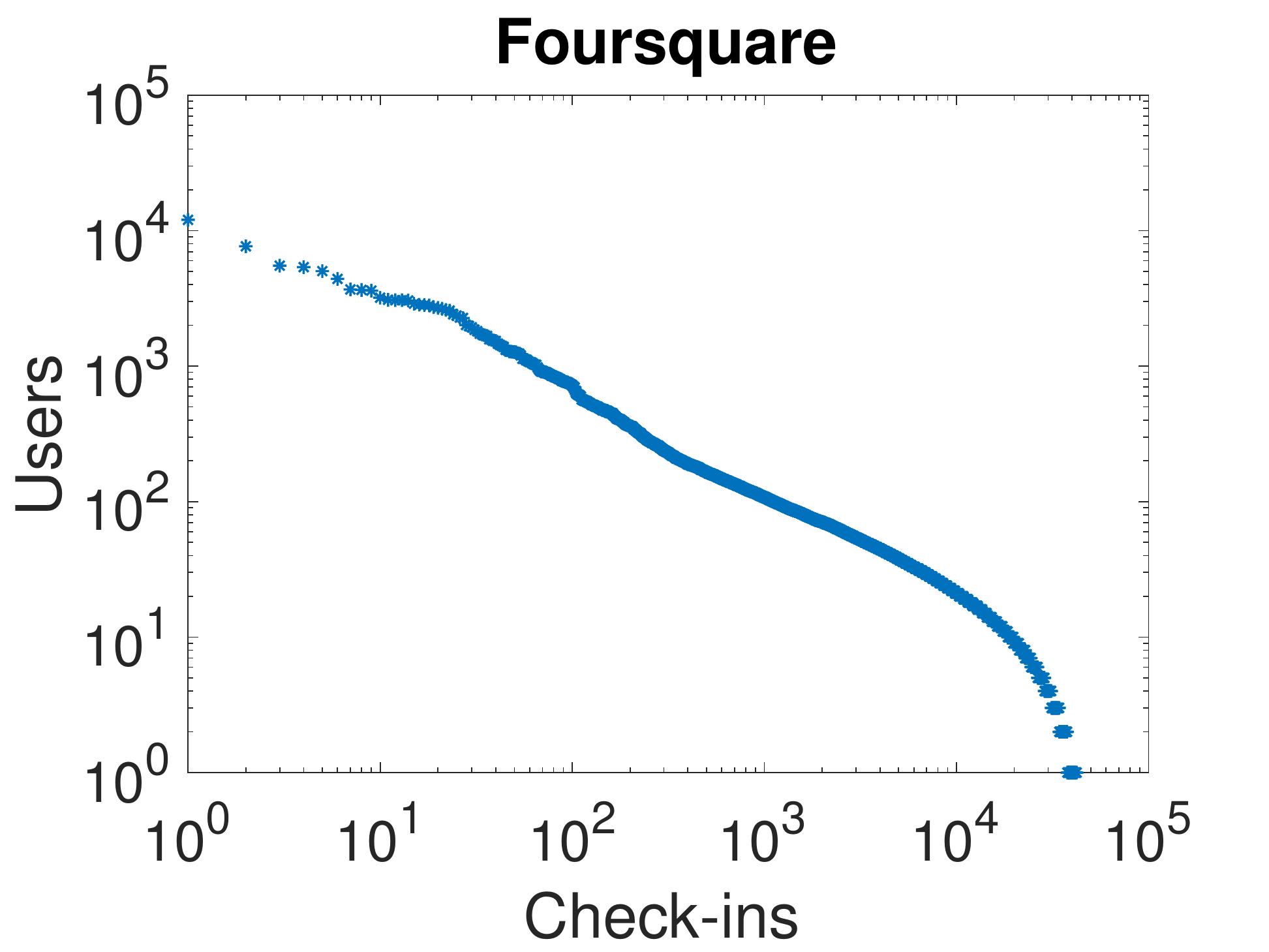}} 
\centerline{(a)}
\end{minipage}
\begin{minipage}{2.9cm}
\centerline{\includegraphics[width=3.1cm]{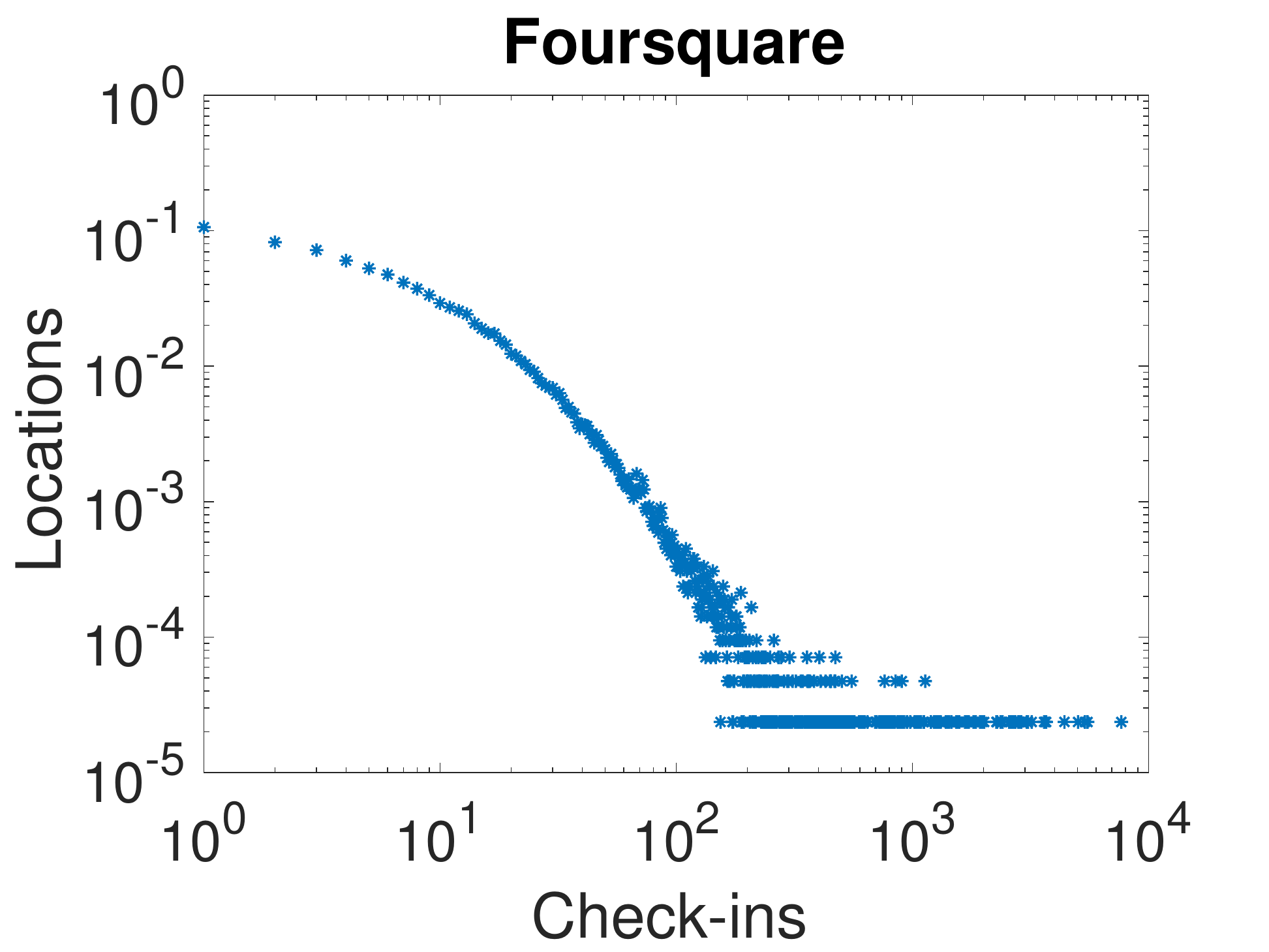}} 
\centerline{(b)}
\end{minipage}
\begin{minipage}{2.9cm}
\centerline{\includegraphics[width=3.1cm]{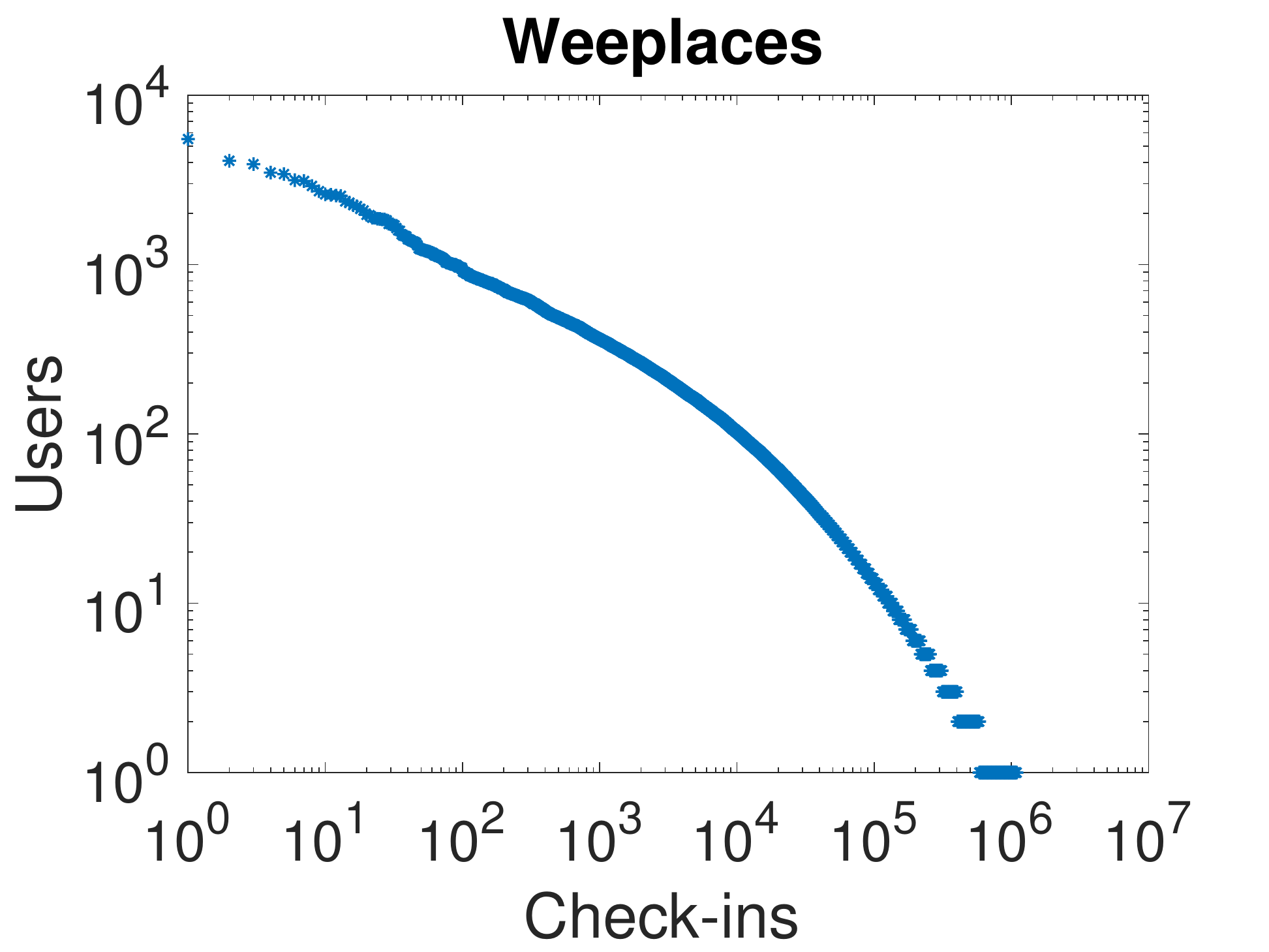}}
\centerline{(c)}
\end{minipage}
\begin{minipage}{2.9cm}
\centerline{\includegraphics[width=3.1cm]{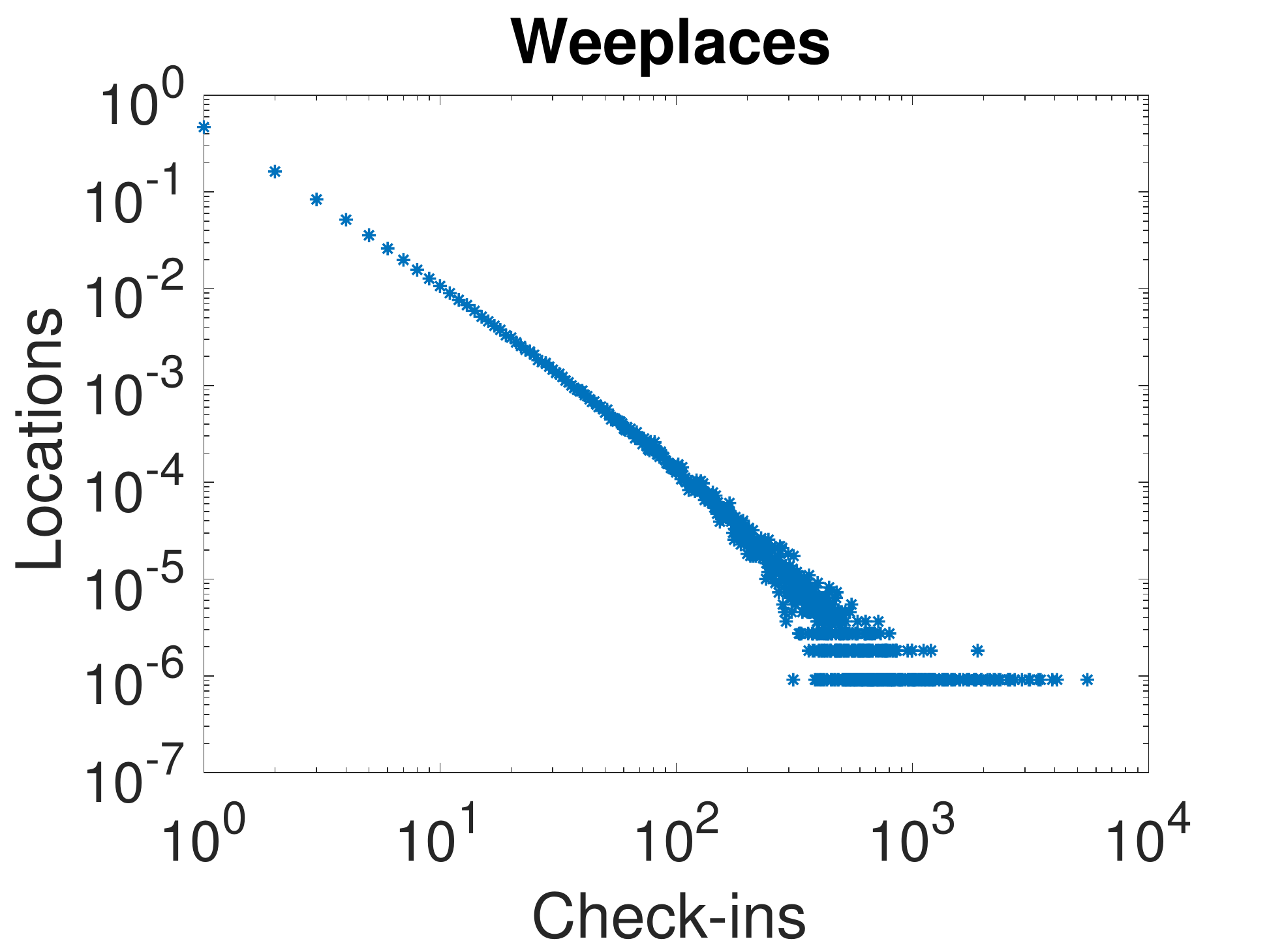}}
\centerline{(d)}
\end{minipage}
\begin{minipage}{2.9cm}
\centerline{\includegraphics[width=3.1cm]{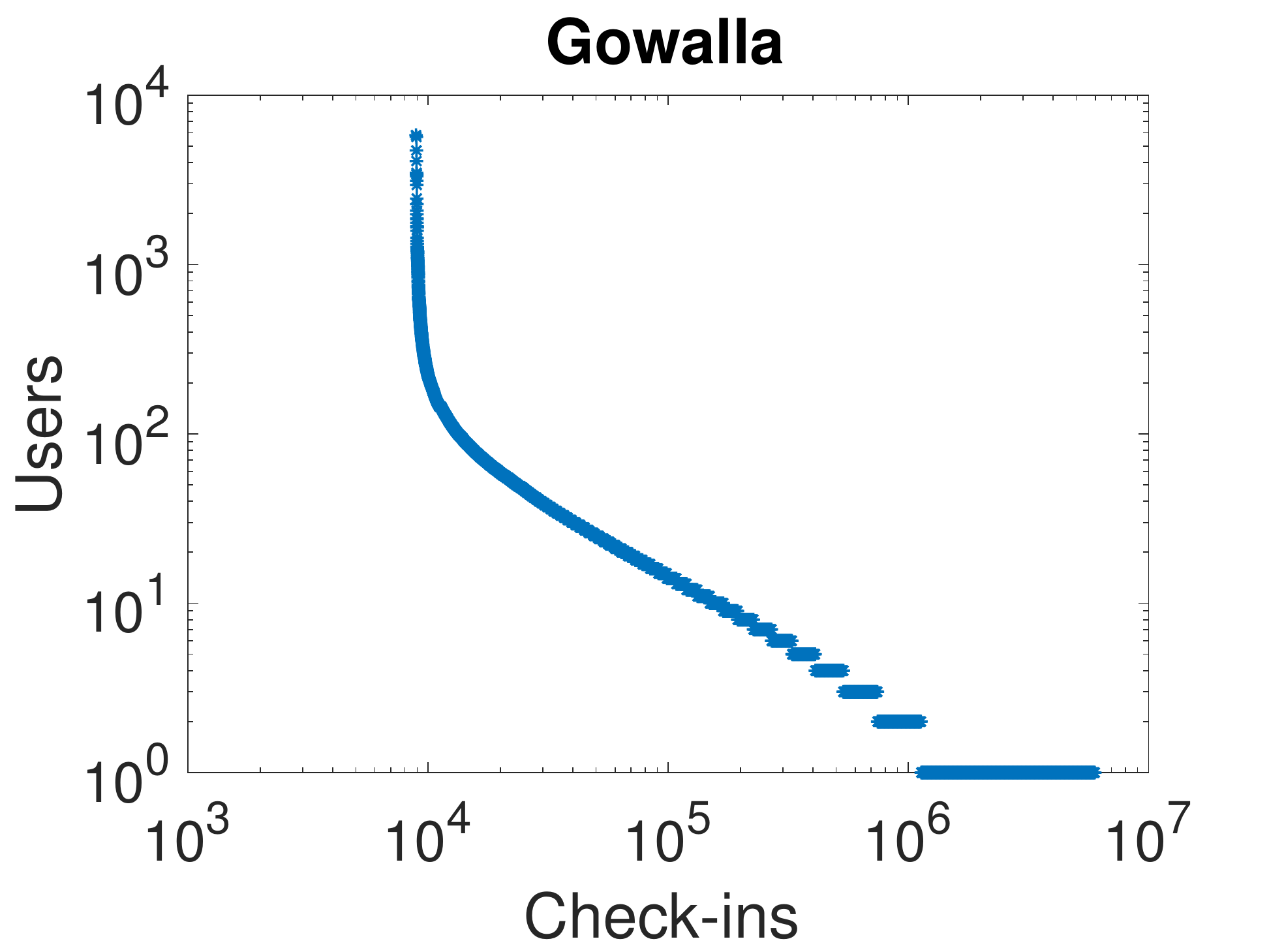}}
\centerline{(e)}
\end{minipage}
\begin{minipage}{2.9cm}
\centerline{\includegraphics[width=3.1cm]{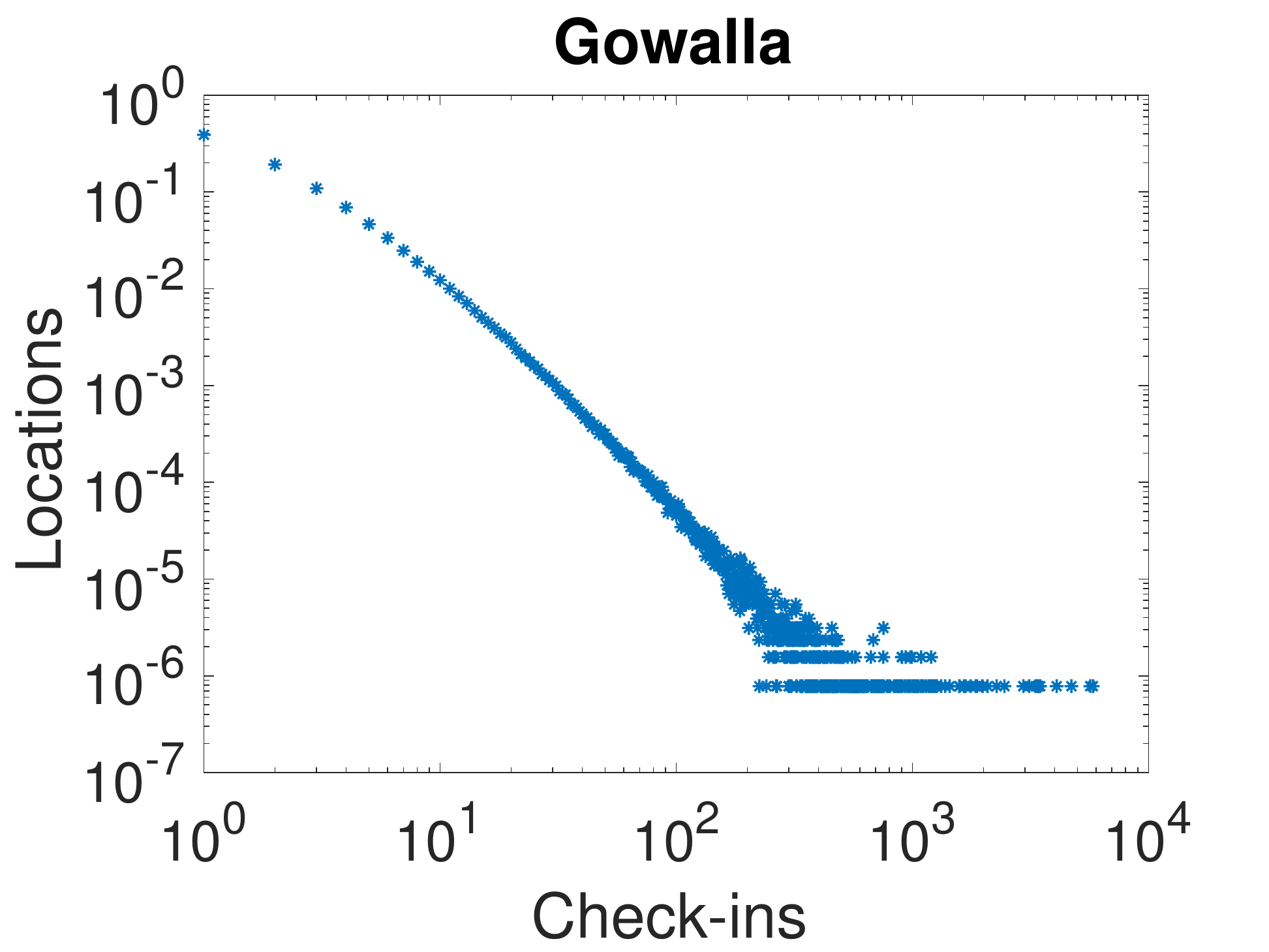}}
\centerline{(f)}
\end{minipage}
\end{center}
\vspace{-0.4cm}
\caption{Power Law distribution diagrams for Foursquare [(a) and (b)], Weeplaces [(c) and (d)], and Gowalla [(e) and (f)].}
\label{fig:Distributions} 
\end{figure*}

\subsection{Datasets and Techniques}
\label{sec:Dataset}

We have used three real-world datasets: $i$) Foursquare~\cite{Xie2016}\footnote{\url{https://sites.google.com/site/dbhongzhi}}, $ii$) Weeplaces~\cite{Baral2016}\footnote{\url{http://www.yongliu.org/datasets}}, and $iii$) Gowalla~\cite{Baral2016}$^5$. Their main characteristics are presented in Table~\ref{tab:Datasets specifications}. All datasets contain users' check-in history with timestamp and geographical information. Additionally, they contain information about the social ties of users' friendships. The datasets span a time period of 5, 91 and 31 months, respectively.

\begin{table}[!ht]
	\caption{Summary of datasets.}
	\label{tab:Datasets specifications}
\centering
\scalebox{0.8}{	
	\begin{tabular}{|c|c|c|c|}\hline
				& Foursquare 		& Weeplaces				& Gowalla   			\\ \hline
	Users   	& 114,508			& 15,799 				&  319,063 				\\ \hline
	POIs		& 62,462			& 971,309 				&  2,844,076	  		\\ \hline
	Check-ins	& 1,434,668			& 7,658,368   			&  36,001,959 			\\ \hline
	Friendships	& 32,511			& 119,930   			&  1,900,653 			\\ \hline
	Time span	& Sep 2010 - Jan 2011 & Nov 2003 - Jun 2011 & Jan 2009 - Aug 2011	\\ \hline
	\end{tabular}}
\end{table}

\input{LaTeX/TimeDistribution.tex}

In Figure~\ref{fig:Distributions} we present the distributions for all three datasets. As shown, the datasets follow a power law distribution for both the number of users' check-ins and the number of visits to a particular location as presented in. According to the power law distribution, there is a small number of users with many check-ins (short head) and many users that have a small check-in record (long tail). Similarly, the same principle holds for locations where, on the one hand, there are popular locations with a lot of visits, and on the other hand, locations that are visited fewer times. Moreover, all datasets present a good example of the cold start problem~\cite{Herlocker2004}, i.e., to recommend new locations to users with small history record.

Furthermore, we present the temporal distribution of all datasets in Figure~\ref{fig:CheckinTemporalDistributions}. In particular, Figure~\ref{fig:CheckinPerDayDistributions} presents the distribution of the check-ins per day of the week. It is noticeable that users tend to be more active from Thursday to Sunday than the rest days of the week. Due to the fact that users check-in locations during their spare time, this figure reflects a tendency according to which they use to perform more check-ins during the end of the week. The same principal stands for the distribution of the check-ins during the day as shown in Figure~\ref{fig:CheckinPerHourDistributions}. Once again, we observe that users are more active during 13:00 p.m. to 2:00 a.m. with peak time at 18:00 p.m. and they tend to check-in at more locations during the night than during the morning.

In the experimentation we evaluate the performance of the following techniques:

\begin{itemize}
  \item {\bf Rank-GeoFM~\cite{Li2015}:} is a MF unified model which learns users' preferences and incorporates the spatial influence of proximate POIs simultaneously. The second term of the model is regularized with a distance-based weighting related to the target POI.
  \item {\bf ASMF~\cite{Li2016}:} is another MF model consisted by a two step procedure. First, the model learns candidate POIs which have been visited by friends (social influence), and then a categorical-based weight is applied considering geographical influence. 
  \item {\bf GE~\cite{Xie2016}:} is a Graph Factorization (GF) approach according to which two joint probability distributions are computed for each pair of nodes. The first is related to the adjacency matrix and the second is related to the embeddings. This method embed four bipartite information networks into the same latent space to predict the next unvisited POI by updating dynamically the users' profile.  
  \item {\bf PACE~\cite{Yang2017}:} is a semi-supervised learning framework, based on users' Preference and Context Embeddings, that jointly learns users' and locations embeddings.
  \item {\bf Versions of RELINE:} To evaluate the influence of each network we have used three versions of our model. In particular, we start with a simplistic version and then we proceed with more enhanced alternatives. 
  		\begin{itemize}
			\item {\bf RELINE$_{V_1}$:} Is a simplified version of RELINE, which contains only the social influence information, that is $(\vv{\textbf{u}}^T   \cdot \vv{\textbf{l}})$.
			\item {\bf RELINE$_{V_2}$:} Is the previous model enriched with the geographical influence $(\vv{\textbf{r}}^T \cdot \vv{\textbf{l}})$.   	
			\item {\bf RELINE$_{V_3}$:} Is the last version which enriches the former model with the Temporal influence $(\vv{\textbf{t}}^T \cdot \vv{\textbf{l}})$.
		\end{itemize}	
\end{itemize}

\begin{table*}[!t]
	\caption{Impact of adding information networks to the model.}
	\label{tab:Impact of information networks}
\centering
\subtable[Foursquare]{
\scalebox{0.74}{
	\begin{tabular}{|l|c|c|c|c|c|}\hline
		\backslashbox{Model}{Acc@$n$}
				& 1			& 5			& 10		& 15		& 20 		\\\hline
RELINE$_{V_1}$  & 0.223		& 0.288 	& 0.334 	& 0.358		& 0.385		\\ \hline
RELINE$_{V_2}$	& 0.257		& 0.291 	& 0.342 	& 0.377		& 0.407 	\\ \hline
RELINE$_{V_3}$ 	& 0.264		& 0.302 	& 0.357 	& 0.391		& 0.409		\\ \hline
RELINE			&{\bf0.286}	&{\bf0.341} &{\bf0.410} &{\bf0.435}	&{\bf0.462}	\\ \hline
	\end{tabular}\label{tab:FoursquareIN}}}
\subtable[Weeplaces]{
\scalebox{0.74}{
	\begin{tabular}{|l|c|c|c|c|c|}\hline
		\backslashbox{Model}{Acc@$n$}
				& 1			& 5			& 10		& 15		& 20 		\\\hline
RELINE$_{V_1}$  & 0.315		& 0.349 	& 0.406 	& 0.420		& 0.445		\\ \hline
RELINE$_{V_2}$ 	& 0.344		& 0.383 	& 0.427 	& 0.458		& 0.479		\\ \hline
RELINE$_{V_3}$	& 0.331		& 0.377 	& 0.413 	& 0.445		& 0.463 	\\ \hline
RELINE			&{\bf0.386}	&{\bf0.421} &{\bf0.488} &{\bf0.514}	&{\bf0.536}	\\ \hline
	\end{tabular}\label{tab:BrightkiteIN}}}
\subtable[Gowalla]{
\scalebox{0.74}{
	\begin{tabular}{|l|c|c|c|c|c|c|}\hline
		\backslashbox{Model}{Acc@$n$}
				& 1			& 5			& 10		& 15		& 20 		\\\hline
RELINE$_{V_1}$  & 0.363		& 0.405 	& 0.505 	& 0.518		& 0.526		\\ \hline
RELINE$_{V_2}$	& 0.355		& 0.401 	& 0.492 	& 0.505		& 0.519 	\\ \hline
RELINE$_{V_3}$ 	& 0.347		& 0.387 	& 0.473 	& 0.493		& 0.508		\\ \hline
RELINE			&{\bf0.408}	&{\bf0.447} &{\bf0.518} &{\bf0.533}	&{\bf0.556}	\\ \hline
	\end{tabular}\label{tab:GowallaIN}}}
\end{table*}

\input{LaTeX/ComparisonAll.tex}

\subsection{Evaluation Methodology}
\label{sec:Evaluation protocol}
We consider the partitioning of check-ins of each target user into three sets: (i) the training set $\cal E_{C}^T$, that is $80\%$ of the total check-ins and is treated as known information, (ii) the probe set $\cal E_{C}^P$, that is $10\%$ and is used for testing our model, and (iii) the validation set $\cal E_{C}^V$ is the rest $10\%$ for tuning the hyper-parameters. It holds that, $\cal E_{C}$=$\cal E_{C}^T \cup \cal E_{C}^P \cup \cal E_{C}^V$ and $\cal E_{C}^T \cap \cal E_{C}^P \cap \cal E_{C}^V$=$\oslash$. Thus, for each target user we generate recommendations based only on the POIs in $\cal E_{C}^T$. 

For the evaluation we measure the Accuracy@$n$ as proposed in~\cite{Wang2016}. In particular, for each $l \in \cal E_{C}^P$ given as a query $Q(u,l,t)$, we compute the prediction score for that $l$ along with all unvisited proximate POIs of the target user with Equation~(\ref{eq:RELINE}). We rank the predicted scores into a list and we select the top@$n$ POIs. If the ground truth $l \in \cal E_{C}^P$ appears in the top@$n$, then we have predicted correctly that location (i.e. True Positive), otherwise our prediction is wrong. To compute the overall accuracy of the top@$n$ we average all predictions test cases as: 

\begin{equation}
	\begin{aligned}
				Accuracy@n = \frac{\text{\#True Positive@$n$}}{\cal E_{C}^P}
		\label{eq:Accuracy}
	\end{aligned}
\end{equation}

\subsection{Impact of Information Networks}
\label{sec:Impact of information networks}

Next, we examine the influence of each information network into the overall predictions accuracy. In particular, we explore how beneficial the enrichment is against the top-$n$ predictions. The participant networks examined in this section are i) the {\it social influence}, ii) the {\it geographical influence}, iii) the {\it temporal influence}, and iv) the {\it users' preference dynamics}. Thus, we compare RELINE with the three models RELINE$_{V_1}$, RELINE$_{V_2}$, and RELINE$_{V_3}$, which are described in section~\ref{sec:Dataset}. It is noticeable that as long as we embed more information networks, the accuracy increases as well, as shown in each of the columns of Tables~\ref{tab:FoursquareIN}-\ref{tab:GowallaIN}. Thus, our model gradually alleviates the sparsity problem since it explores more information about the users or the POIs. Moreover, the accuracy of each model increases with $n$, meaning that the models fit well to users' behavior.

\subsection{Comparison with Other Techniques}
\label{sec:Comparison medels}
In this section we compare our model with other state-of-the-art approaches in terms of accuracy for the top-$n$ predictions [$n$=$1,5,10,15,20$] against 3 big datasets. In particular, we examine the performance of all models for providing POI recommendations for $i$) all users, $i$i) cold-start users, and $iii$) cold-start POIs. 

\subsubsection{Accuracy over all users}
\label{sec:AccuracyCSA}
First, we examine how models perform while providing recommendations to all users, without taking into account the history size or data sparsity. It is observed that RELINE significantly outperforms all methods as shown in Figure~\ref{fig:AccuracyALL}. Compared to methods that either a) learn users' preferences and then incorporate geographical influence like Rank-GeoFM, or b) learn social network check-ins like ASMF and ignore i) the sequence of the POIs, ii) the temporal influence, and iii) preference dynamics, the performance is much higher since our method explores richer information in the same latent space. On the other hand, methods that learn users' and POI embeddings into the same latent space through multiple information networks, like GE and PACE, miss other important factors such as i) periodicity, ii) preference evolution, and iii) sequential importance of the check-ins and gain lower accuracy compared to our method. We highlight that all models show higher accuracy while the number of POIs is small and the check-in activity denser, as show in Figure~\ref{fig:AccuracyFoursquareAll}. Moreover, when the dataset is sparse with multiple available POIs to visit, such as the other two datasets presented in Figures~\ref{fig:AccuracyWeeplacesAll}-\ref{fig:AccuracyGowallaAll}, the accuracy is lower. This finding supports the claim that learning both users' and POIs embeddings derived from as many information networks as possible, increases the model's ability to correlate a user with POIs and eventually improves the accuracy.

\subsubsection{Accuracy over the cold-start Users}
\label{sec:AccuracyCSU}
Next, we examine the effectiveness of our model regarding the cold-start user and perform comparisons with other approaches. The concept was initially introduced in~\cite{Herlocker2004} and refers to users with short history. It is evident that supporting recommendations for such users is a difficult task due to lack of adequate information. To this context, we performed experiments providing recommendations only to cold-start users and we compared the performance of the models in terms of accuracy, as shown in Figure~\ref{fig:DistributionsCSU}. Since only ASMF and GE support cold-start recommendations, we compare our model to these two methods. Moreover, we observe that the overall accuracy compared to the experiments of the previous section is reduced, which is normal, since we do not have much information. Even though, ASMF learns the POIs visited by target user's social network and then it refines the results with a categorical weighted strategy w.r.t. geographical influence, its accuracy is significantly lower compared to our method. Similarly, the performance of GE which jointly captures the spatial influence, the sequential effect, the periodicity, and the semantics into the same latent space is lower since it does not consider: 1) the importance of stay points in each sequence, 2) the preference dynamics, 3) the temporal effect, and 4) the social influence. In contrast, we use side information related to both users and POIs from eight weighted information networks, which has a significant impact in effectiveness.

\input{LaTeX/ComparisonCSU.tex}

\input{LaTeX/ComparisonCSP.tex}

\begin{table*}[!t]
	\caption{Impact of parameters $S$ and $d$ for top@$10$}
	\label{tab:Impact of S and d parameters}
\centering
\subtable[Foursquare]{
\scalebox{0.75}{
	\begin{tabular}{|c|c|c|c|c|c|c|}\hline
		\multirow{2}{*}{S(mil)} & \multicolumn{6}{c|}{$d$}	 \\ \cline{2-7}
				& 70	& 80	& 90	&{\bf100}	&110 	& 120	\\\hline
	50   		& 0.383	& 0.385 & 0.385 & 0.388		& 0.388	& 0.388	\\ \hline
	100			& 0.392	& 0.398 & 0.390 & 0.391		& 0.391 & 0.392	\\ \hline
{\bf150} 		& 0.405	& 0.408 & 0.409 & {\bf0.410}& 0.411	& 0.411	\\ \hline
	200			& 0.405	& 0.408 & 0.408 & 0.409		& 0.411	& 0.411	\\ \hline
	250			& 0.423	& 0.408 & 0.408 & 0.409		& 0.411	& 0.411 \\ \hline
	\end{tabular}\label{tab:FoursquareSD}}}
~
\subtable[Weeplaces]{
\scalebox{0.75}{
	\begin{tabular}{|c|c|c|c|c|c|c|}\hline
		\multirow{2}{*}{S(mil)} & \multicolumn{6}{c|}{$d$}	 \\ \cline{2-7}
				& 70	& 80	& 90	&{\bf100}	&110 	& 120	\\\hline
	100   		& 0.464	& 0.471 & 0.474 & 0.476		& 0.476	& 0.476	\\ \hline
	150			& 0.472	& 0.475 & 0.479 & 0.479		& 0.479	& 0.480	\\ \hline
{\bf200} 		& 0.478	& 0.480 & 0.484 &{\bf0.488}	& 0.488	& 0.489	\\ \hline
	250			& 0.478	& 0.481 & 0.485 & 0.488		& 0.489	& 0.489	\\ \hline
	300			& 0.479	& 0.481 & 0.485 & 0.489		& 0.489	& 0.490	\\ \hline
	\end{tabular}\label{tab:BrightkiteSD}}}
~
\subtable[Gowalla]{
\scalebox{0.75}{
	\begin{tabular}{|c|c|c|c|c|c|c|}\hline
		\multirow{2}{*}{S(mil)} & \multicolumn{6}{c|}{$d$}	 \\ \cline{2-7}
				& 110	& 120	& 130	&{\bf140}	&150 	& 160	\\\hline
	200   		& 0.489	& 0.494 & 0.498 & 0.502		& 0.502	& 0.503	\\ \hline
	250			& 0.498	& 0.502 & 0.507 & 0.510		& 0.510	& 0.511	\\ \hline
{\bf300} 		& 0.503	& 0.509 & 0.514 &{\bf0.518}	& 0.518	& 0.519	\\ \hline
	350			& 0.503	& 0.510 & 0.515 & 0.518		& 0.519	& 0.519	\\ \hline
	400			& 0.504	& 0.510 & 0.515 & 0.519		& 0.520	& 0.520	\\ \hline
	\end{tabular}\label{tab:GowallaSD}}}
\end{table*}

\subsubsection{Accuracy over cold-start POIs.}
\label{sec:AccuracyCSP}
Next, we examine a similar problem called cold-start POI. The goal in this case is to recommend unvisited POIs to users that have at least one check-in at a POI that has less than 15 check-ins. Thus, we examine not only how models behave on new users but also, how they behave when a new location gets into the system. In simple terms, we want to check if the new location is among the top-$n$ recommendations or not.

Once again, we evaluate our approach only with the models that support cold-start POI problem. In Figure~\ref{fig:DistributionsCSP}, we present the results of all models against all three datasets. Clearly, our model outperforms compared methods since, among all aforementioned factors, it explores POIs as a sequence of routes. In particular, all cold-start POIs are correlated with other proximate POIs when we extract users' routes. Also, some of them are considered as `stay points' based on how long a user spends on that location. Then, we weigh the edge between the user and the cold-start POI, as an important one, if that user spends a lot of time there. This way, we tackle cold-start POIs by using their relational influence with other nodes on the graph during learning phase. It is noticeable that both comparison models gain lower accuracy for top-$1$ prediction when the number of the POIs is small as shown in Figures~\ref{fig:AccuracyFoursquareCSP}-\ref{fig:AccuracyWeeplacesCSP} because they either explore users' friends check-ins who have not visited cold-start POIs, or they explore POIs sequences based on the frequency a user check-ins a cold-start POI. The POIs with few check-ins are not connected to others in the POI-POI graph which GE uses. Finally, both comparison models seem to gain higher accuracy with many POIs on the system as for example the Gowalla dataset which is pointed in Figure~\ref{fig:AccuracyGowallaCSP}.

\begin{table*}[!t]
	\caption{Impact of time interval $\Delta$T over accuracy for the top@$n$ recommendations}
	\label{tab:Impact of of time interval}
\centering
\subtable[Foursquare]{
\scalebox{0.9}{
	\begin{tabular}{|c|c|c|c|c|c|} \hline
	\multirow{2}{*}{$\Delta$T} & \multicolumn{5}{c|}{Acc@$n$}	 \\ \cline{2-6}
				& 1			& 5			& 10		& 15		& 20 		\\\hline
	5			& 0.254		& 0.317 	& 0.381 	& 0.396	 	& 0.438		\\ \hline
	10 			& 0.261		& 0.325 	& 0.397 	& 0.403		& 0.446		\\ \hline
	15			& 0.273		& 0.328  	& 0.402  	& 0.425		& 0.453		\\ \hline
{\bf20}			&{\bf0.286}	&{\bf0.341} &{\bf0.410} &{\bf0.435}	&{\bf0.462}	\\ \hline
	25 			& 0.273		& 0.335   	& 0.402   	& 0.420		& 0.453		\\ \hline
	30 			& 0.265		& 0.328  	& 0.397  	& 0.416		& 0.446		\\ \hline
	\end{tabular}\label{tab:FoursquareTT}}}
~
\subtable[Weeplaces]{
\scalebox{0.9}{
	\begin{tabular}{|c|c|c|c|c|c|c|}\hline
	\multirow{2}{*}{$\Delta$T} & \multicolumn{5}{c|}{Acc@$n$}	 \\ \cline{2-6}
				& 1			& 5			& 10		& 15		& 20 		\\ \hline
	10   		& 0.356		& 0.398		& 0.457 	& 0.489		& 0.510		\\ \hline
	20			& 0.368		& 0.405 	& 0.463		& 0.497		& 0.519		\\ \hline
	30 			& 0.373		& 0.413 	& 0.475		& 0.501		& 0.524		\\ \hline
{\bf40}			&{\bf0.386}	&{\bf0.421} &{\bf0.488} &{\bf0.514}	&{\bf0.536}	\\ \hline
	50			& 0.374		& 0.417  	& 0.479  	& 0.507		& 0.527		\\ \hline
	60 			& 0.366		& 0.404		& 0.463		& 0.498		& 0.516		\\ \hline
	\end{tabular}\label{tab:BrightkiteTT}}}
~
\subtable[Gowalla]{
\scalebox{0.9}{
	\begin{tabular}{|c|c|c|c|c|c|c|}\hline
     \multirow{2}{*}{$\Delta$T} & \multicolumn{5}{c|}{Acc@$n$}	 \\ \cline{2-6}
				& 1			& 5			& 10		& 15		& 20 		\\\hline
	1 			& 0.381		& 0.421  	& 0.490 	& 0.509		& 0.528		\\ \hline
	5   		& 0.389		& 0.428  	& 0.498  	& 0.517		& 0.536		\\ \hline
	10			& 0.397		& 0.436  	& 0.507  	& 0.524		& 0.545		\\ \hline
{\bf15} 		&{\bf0.408}	&{\bf0.447} &{\bf0.518} &{\bf0.533}	&{\bf0.556}	\\ \hline
	20			& 0.391 	& 0.433		& 0.504  	& 0.526		& 0.542		\\ \hline
	30			& 0.379		& 0.427		& 0.495  	& 0.515		& 0.537		\\ \hline
	\end{tabular}\label{tab:GowallaTT}}}
\end{table*}

\input{LaTeX/TunningParameters.tex}

\subsection{Parameter Tuning}
\label{sec:Parameters Tuning}
In this section, we study the importance of parameter tuning. In particular, we examine the impact of: 1) adding information networks to the model, 2) the number of samples $S$, 3) the embeddings dimensionality $d$, and 4) the time period size $T$, to the performance of our model in terms of accuracy.

\subsubsection{Impact of Samples and Dimensions number}
\label{sec:INSD}

Here, we present the experiments conducted to select the best candidate parameters for the number of samples and dimensions. The results for each dataset are presented in Table~\ref{tab:Impact of S and d parameters}. Our findings for the top@10 indicate that our model is not greatly affected by the dimensionality $d$. The accuracy increases with higher rate along with dimensionality until $d$=100 for Foursquare and Weeplaces and $d$=140 for Gowalla. Thereafter, the accuracy does not change significantly. In contrast to dimensionality, our model is sensitive to the number of the samples ($S$). Until a convergence point is reached, our model keeps increasing its accuracy along with the size of $S$ and then the improvement is poor. Also, the higher the number of connections between the network edges, the higher the number of the sample. To gain higher accuracy, we set the number of samples equal to 100 for Foursquare, 200 for Weeplaces, and 300 for Gowalla along with the dimensionalities discussed previously. 

\subsubsection{Impact of Time Period}
\label{sec:ITP}

Next, we examine the influence of the time interval $\Delta$T to the overall accuracy against varying values of the top@$n$ predictions. $\Delta$T is crucial for our model since it is used to construct multiple graphs such as User-Route, User-Time period, POI-Time period etc. When extracting user-route edges, if there are not enough check-in data during small time intervals, the correlation of that user with other candidate POIs is difficult to be achieved. To overcome this issue, we use different size for $\Delta$T to examine which achieves higher accuracy. Table~\ref{tab:Impact of of time interval} presents the results for each dataset separately. It is noticeable that there is a point in each sub-table where the accuracy reaches its maximum value and then gradually decreases. The reason is that, when $\Delta$T is too small, there are few data and the accuracy is low. On the other hand, when $\Delta$T is large, there are too many nodes correlated to the target user which leads to overfitting. Thus, we set the size equal to 20, 40, and 15 for Foursquare, Weeplaces, and Gowalla, respectively.

\subsubsection{Tuning parameters $\alpha, \beta, \gamma$, and $\delta$}
\label{sec:TPABGD}

In Figure~\ref{fig:Tuning parameters}, we present the results of tuning the parameters $\alpha, \beta, \gamma$, and $\delta$ in terms of accuracy for all datasets. As shown in equation~(\ref{eq:RELINE}), each parameter corresponds to the social, geographical and temporal influences, along with the preference dynamics, respectively. For simplicity, we set each parameter $p$ to a value between [0-1] and all other parameters equal to $(1-p)/3$. It is observed that there is an intersection point in each diagram where the accuracy meets. Also, there is a peak point where our model gets the higher accuracy, which was used to train our model. Moreover, the importance of each influential network is diversified trying to adapt the model to users' behavior.

As shown in Figure~\ref{fig:Tuning parameters}, the influential parameter may be different. For example, regarding Foursquare, users are highly influenced by their social network as shown in Figure~\ref{fig:FoursquareParameters}. Thus, they tend to visit the locations their friends visit. On the other hand, Weeplaces users tend to follow specific routes everyday, thus their movement to a new location is highly influenced by geographical factors, as shown in Figure~\ref{fig:WeeplacesParameters}. Finally, in datasets with many check-ins, like Gowalla, it is clear that capturing the preference dynamics influence is more important when predicting the next location as Figure~\ref{fig:GowallaParameters} depicts.

%% file: LaTeX/TimeDistribution.tex
\begin{figure}[!b]
	\begin{subfigure}[]{
\includegraphics[height=3.5cm, width=4cm]{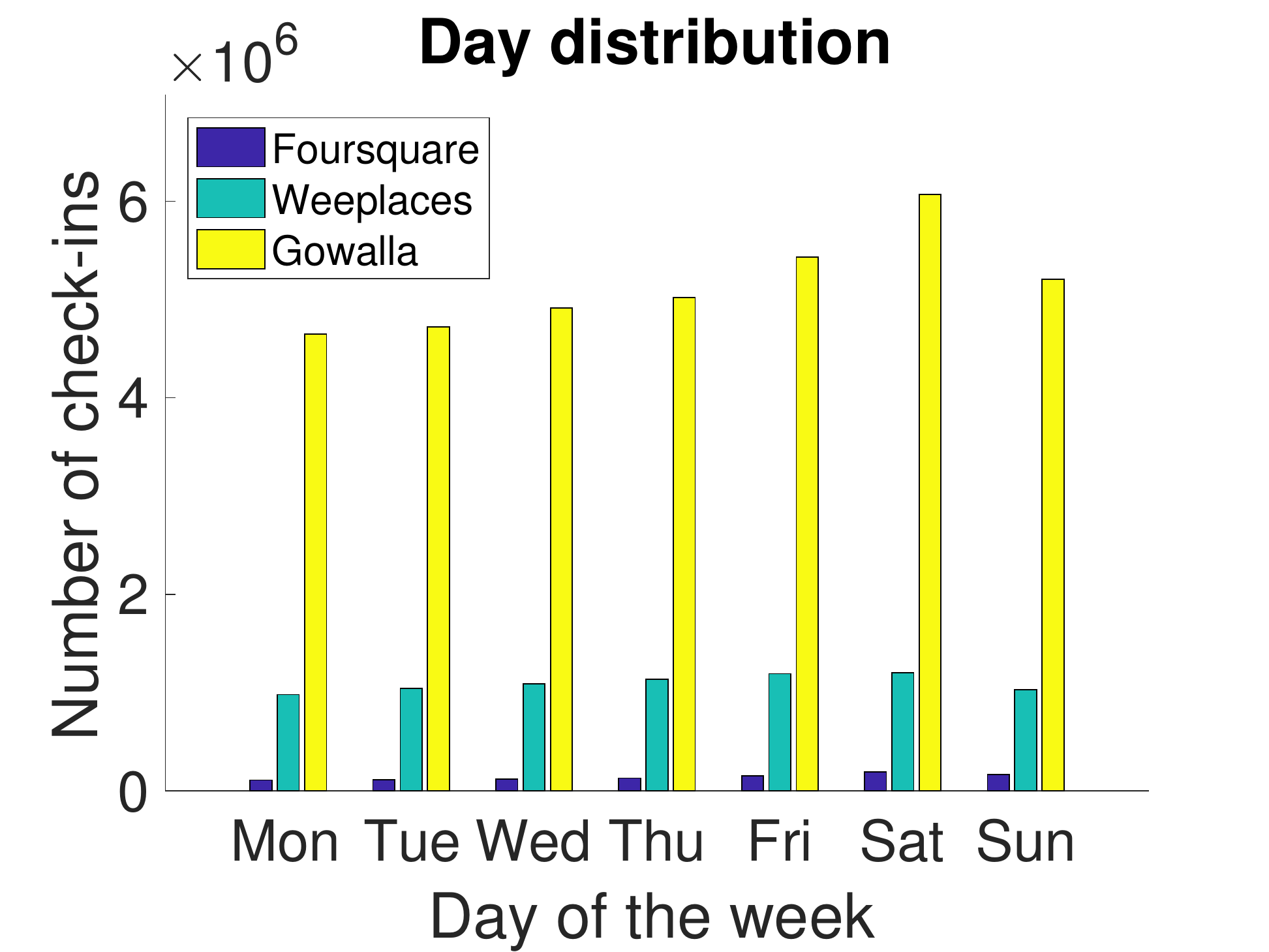}
		\label{fig:CheckinPerDayDistributions} 
		}
	\end{subfigure}
	\begin{subfigure}[]{
\hspace{-0.9cm}			
\includegraphics[height=3.5cm, width=5.3cm]{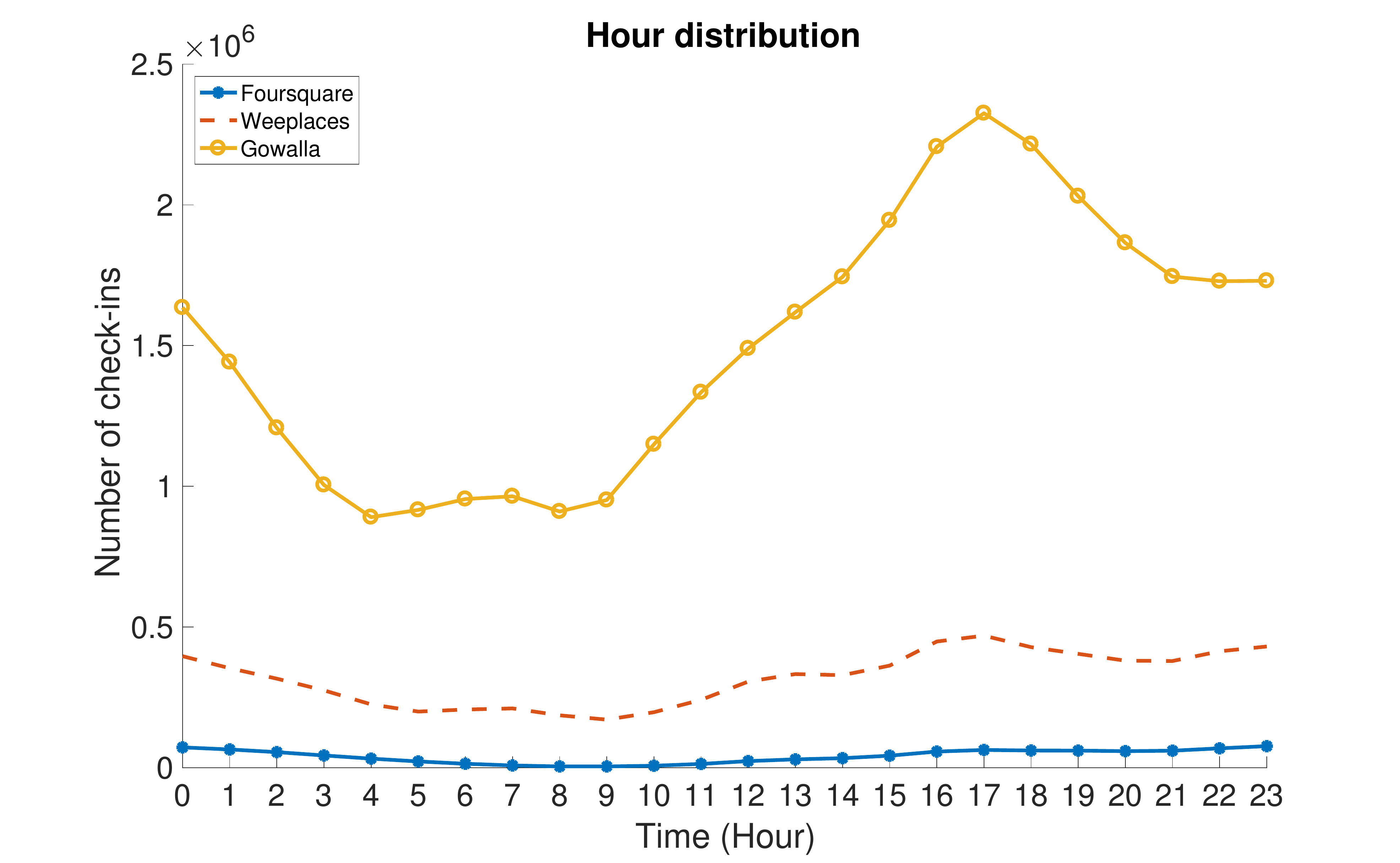}
		\label{fig:CheckinPerHourDistributions} 	
		}
	\end{subfigure}
\vspace{-0.4cm}	
\caption{Check-in distribution per: (a) day of the week, and (b) hour of the day.} 
\label{fig:CheckinTemporalDistributions} 
\end{figure}

%% file: LaTeX/ComparisonAll.tex
\begin{figure*}[!ht]
\centering
	\begin{subfigure}[]{
	\centering			
\includegraphics[height=4.2cm, width=5.7cm]{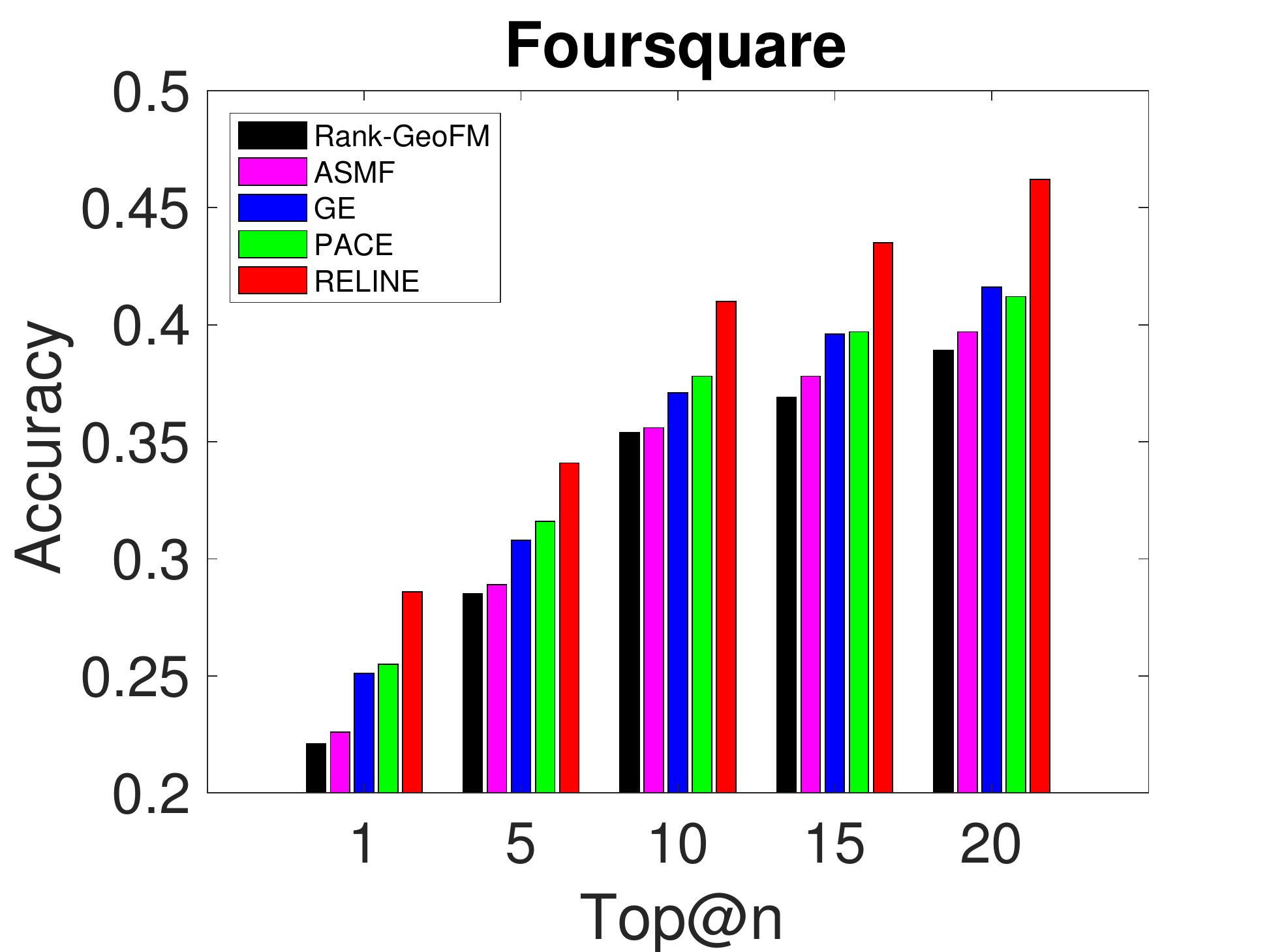}
		\label{fig:AccuracyFoursquareAll} 
		}
	\end{subfigure}
	\begin{subfigure}[]{
	\centering			
\includegraphics[height=4.2cm, width=5.7cm]{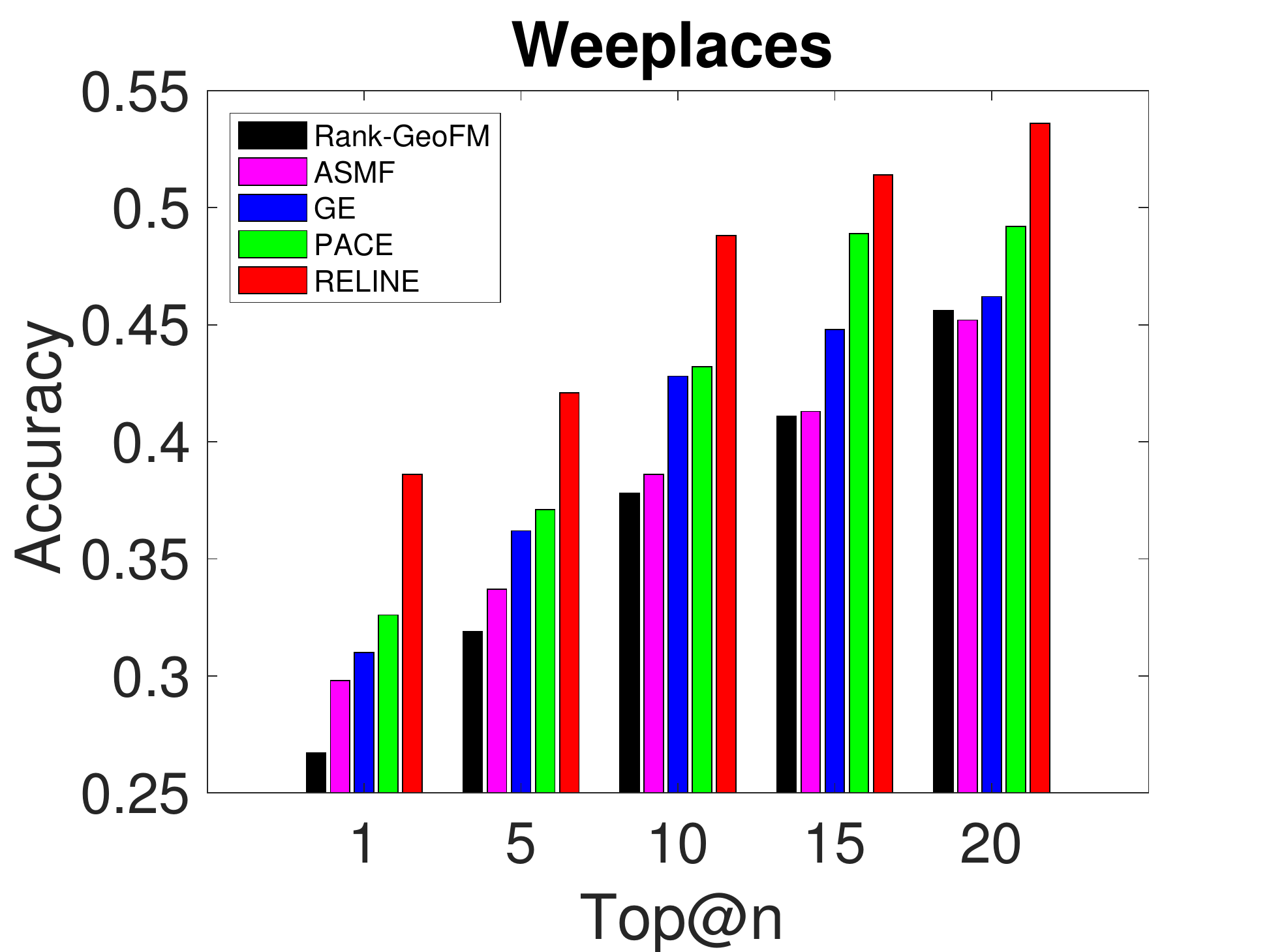}
		\label{fig:AccuracyWeeplacesAll} 
		}
	\end{subfigure}	
	\begin{subfigure}[]{
	\centering	
\includegraphics[height=4.2cm, width=5.7cm]{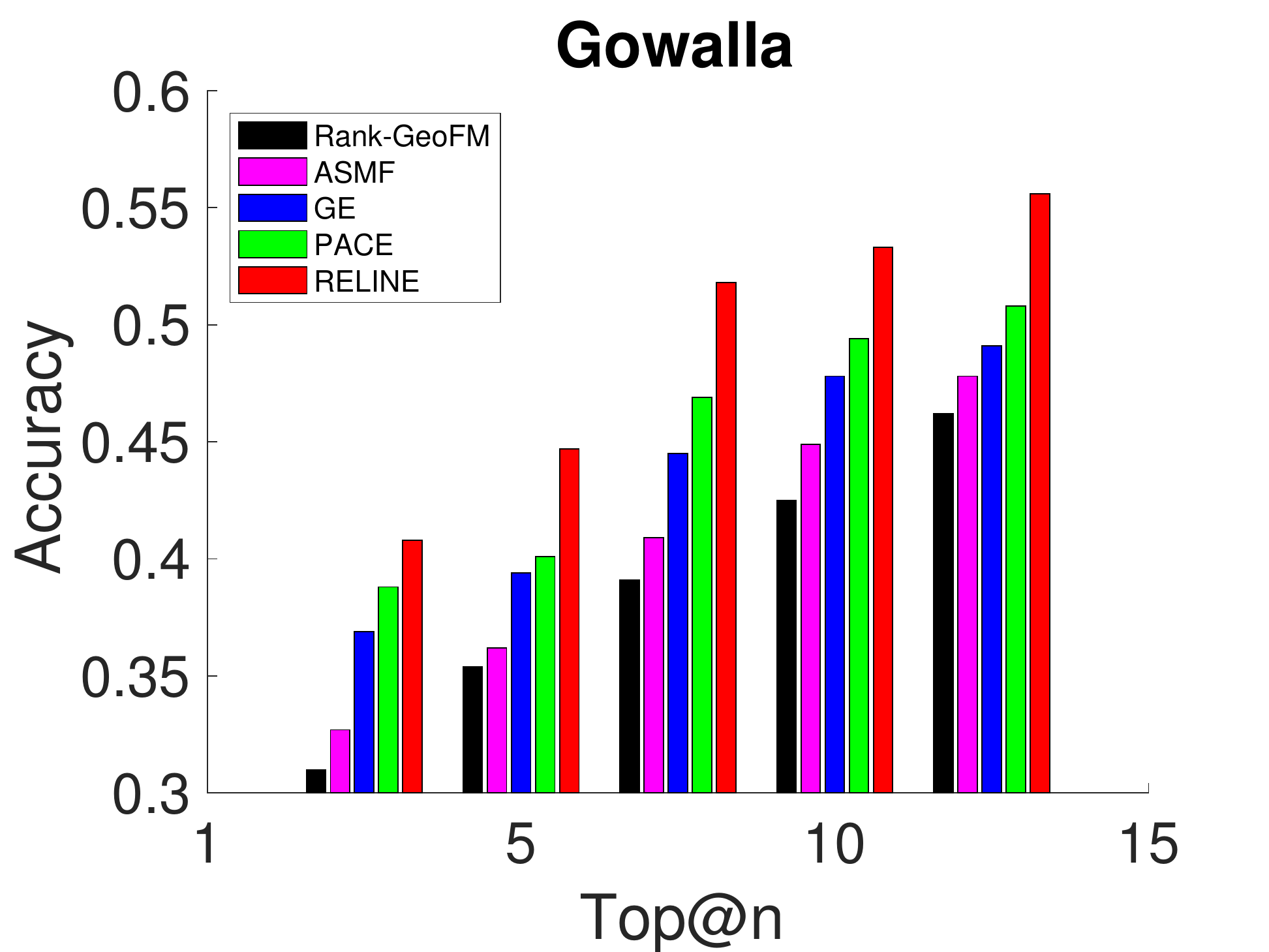}	
		\label{fig:AccuracyGowallaAll}	  
		}
	\end{subfigure}
\vspace{-0.4cm}	
\caption{Accuracy vs the top@$n$ recommendations for all users.} 
\label{fig:AccuracyALL} 
\end{figure*}

%% file: LaTeX/ComparisonCSU.tex
\begin{figure*}[!ht]
\centering
	\begin{subfigure}[]{
	\centering			
\includegraphics[height=4.2cm, width=5.7cm]{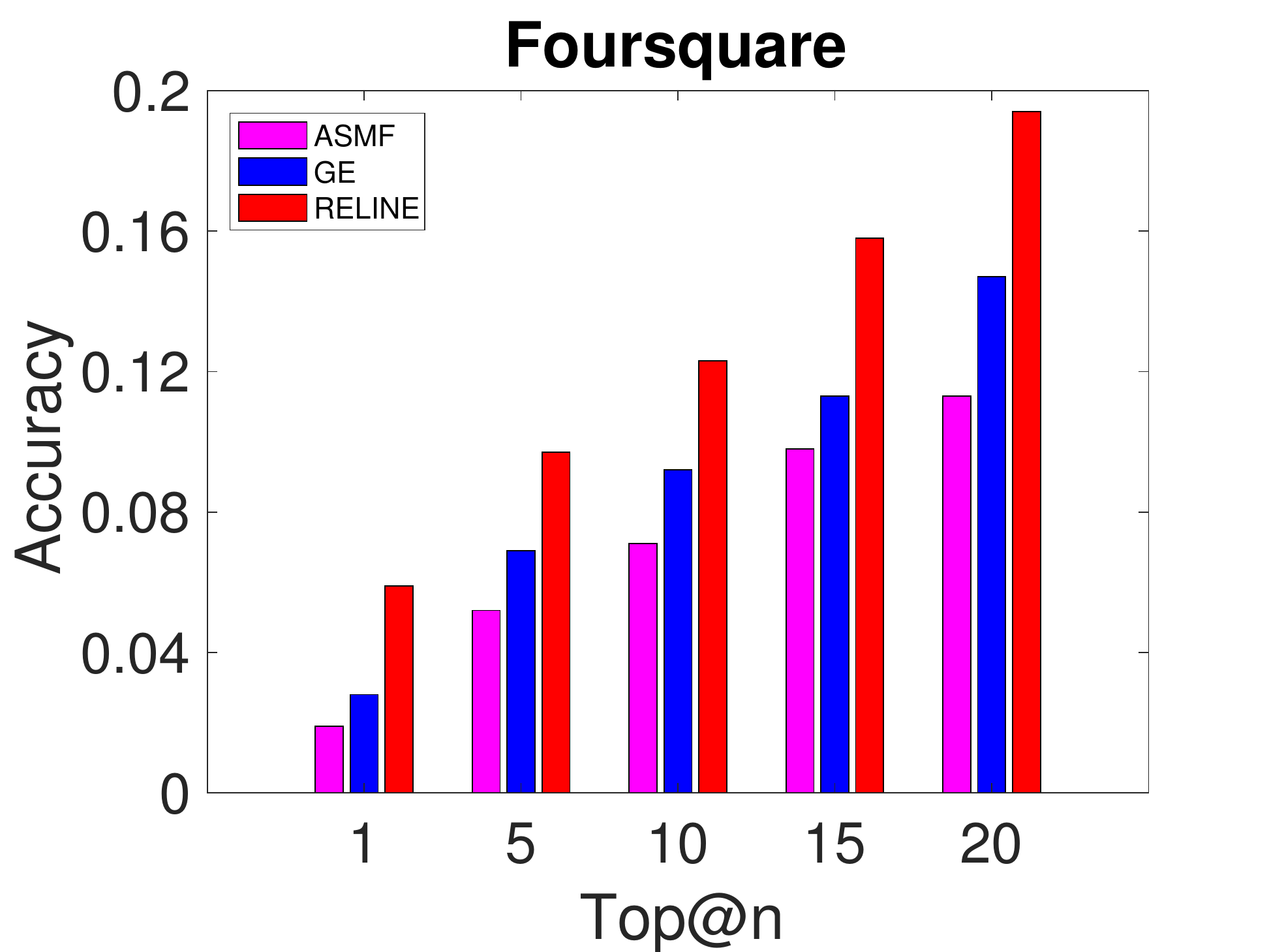}
		\label{fig:AccuracyFoursquareCSU} 
		}
	\end{subfigure}
	\begin{subfigure}[]{
	\centering			
\includegraphics[height=4.2cm, width=5.7cm]{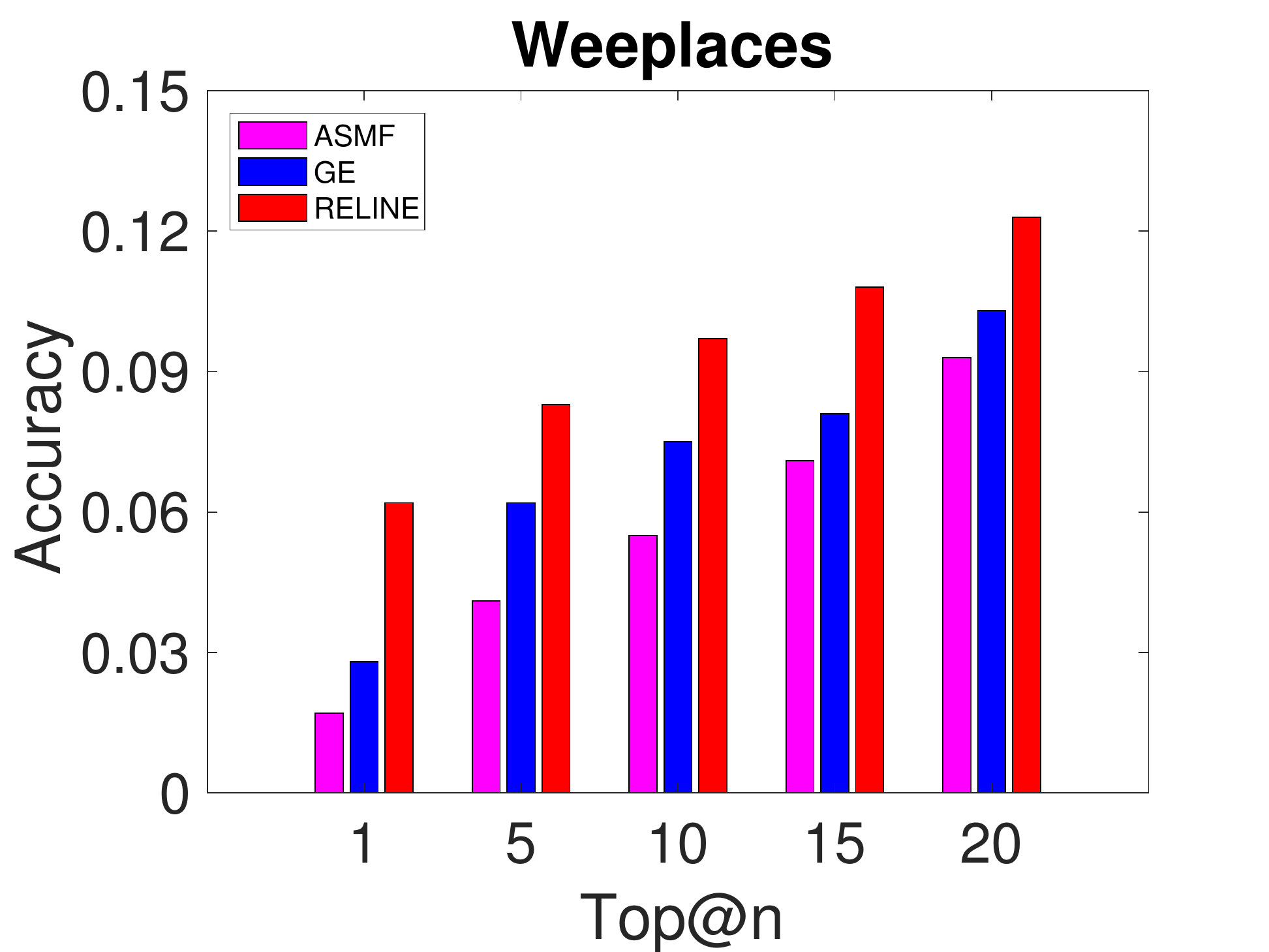}
		\label{fig:AccuracyWeeplacesCSU} 
		}
	\end{subfigure}	
	\begin{subfigure}[]{
	\centering	
\includegraphics[height=4.2cm, width=5.7cm]{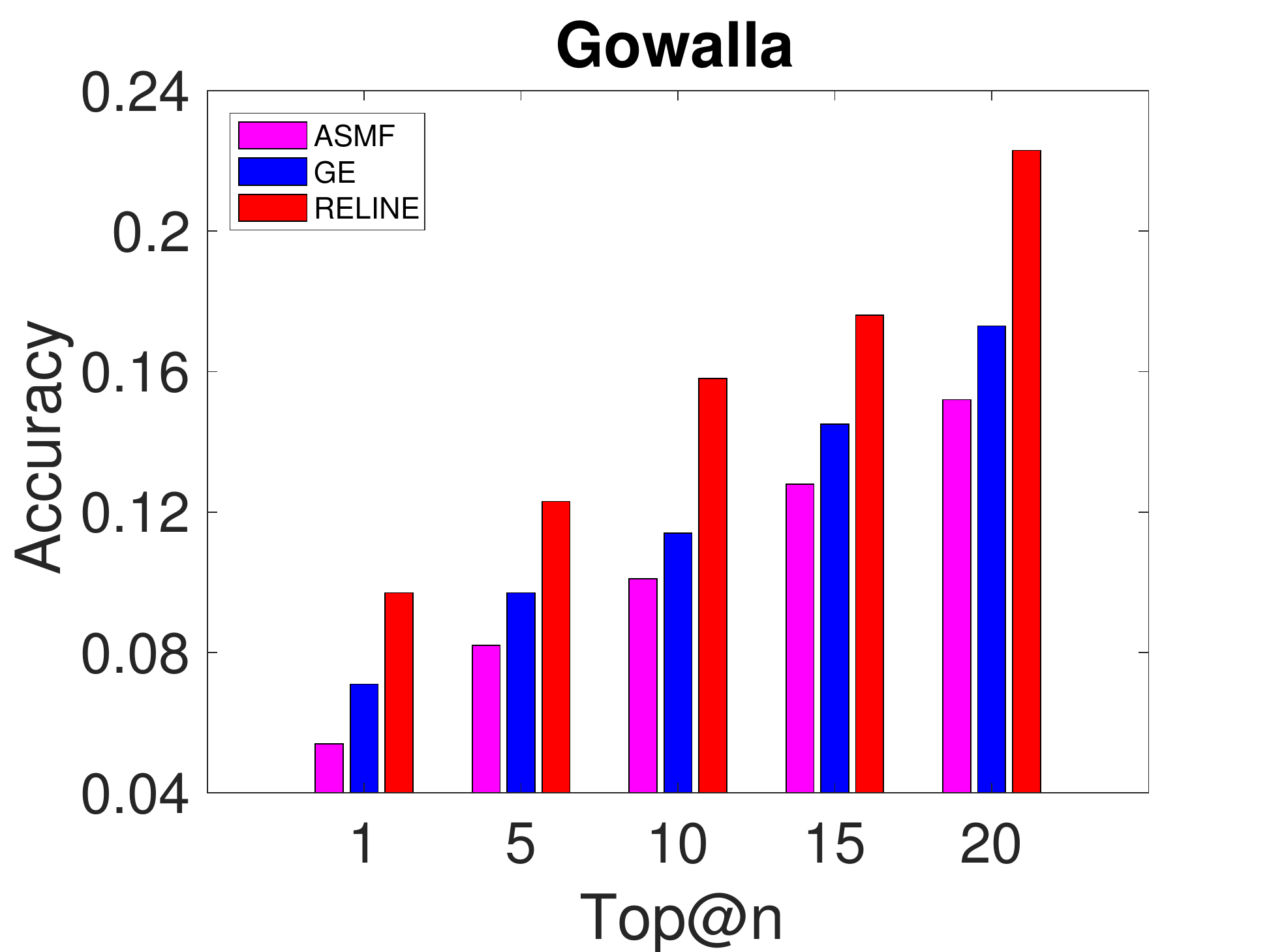}	
		\label{fig:AccuracyGowallaCSU}	  
		}
	\end{subfigure}
\vspace{-0.4cm}
\caption{Accuracy vs the top@$n$ recommendations for cold-start users.} 
\label{fig:DistributionsCSU} 
\end{figure*}

%% file: LaTeX/ComparisonCSP.tex
\begin{figure*}[!ht]
\centering
	\begin{subfigure}[]{
	\centering			
\includegraphics[height=4.2cm, width=5.7cm]{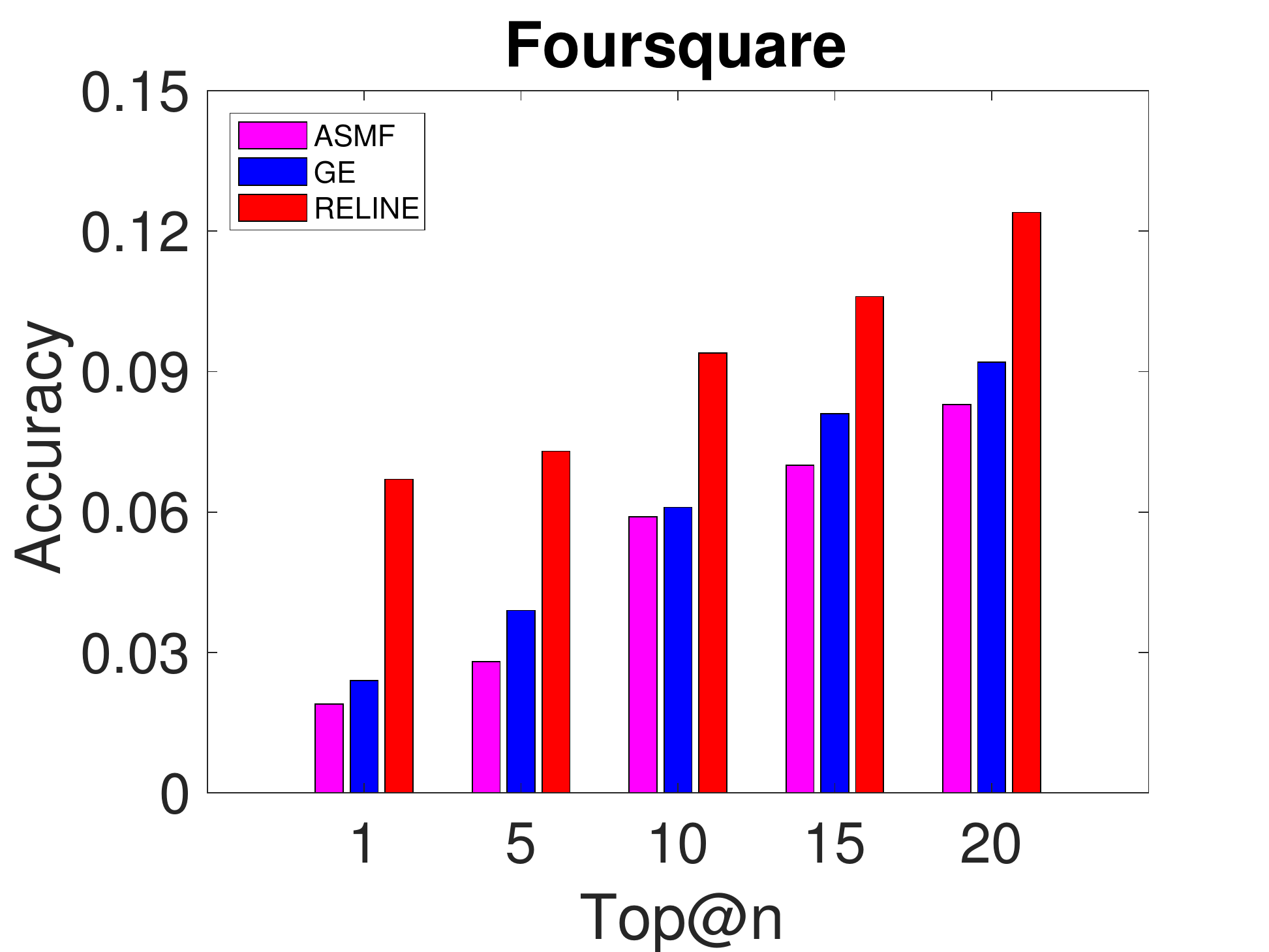}
		\label{fig:AccuracyFoursquareCSP} 
		}
	\end{subfigure}	
	\begin{subfigure}[]{
	\centering			
\includegraphics[height=4.2cm, width=5.7cm]{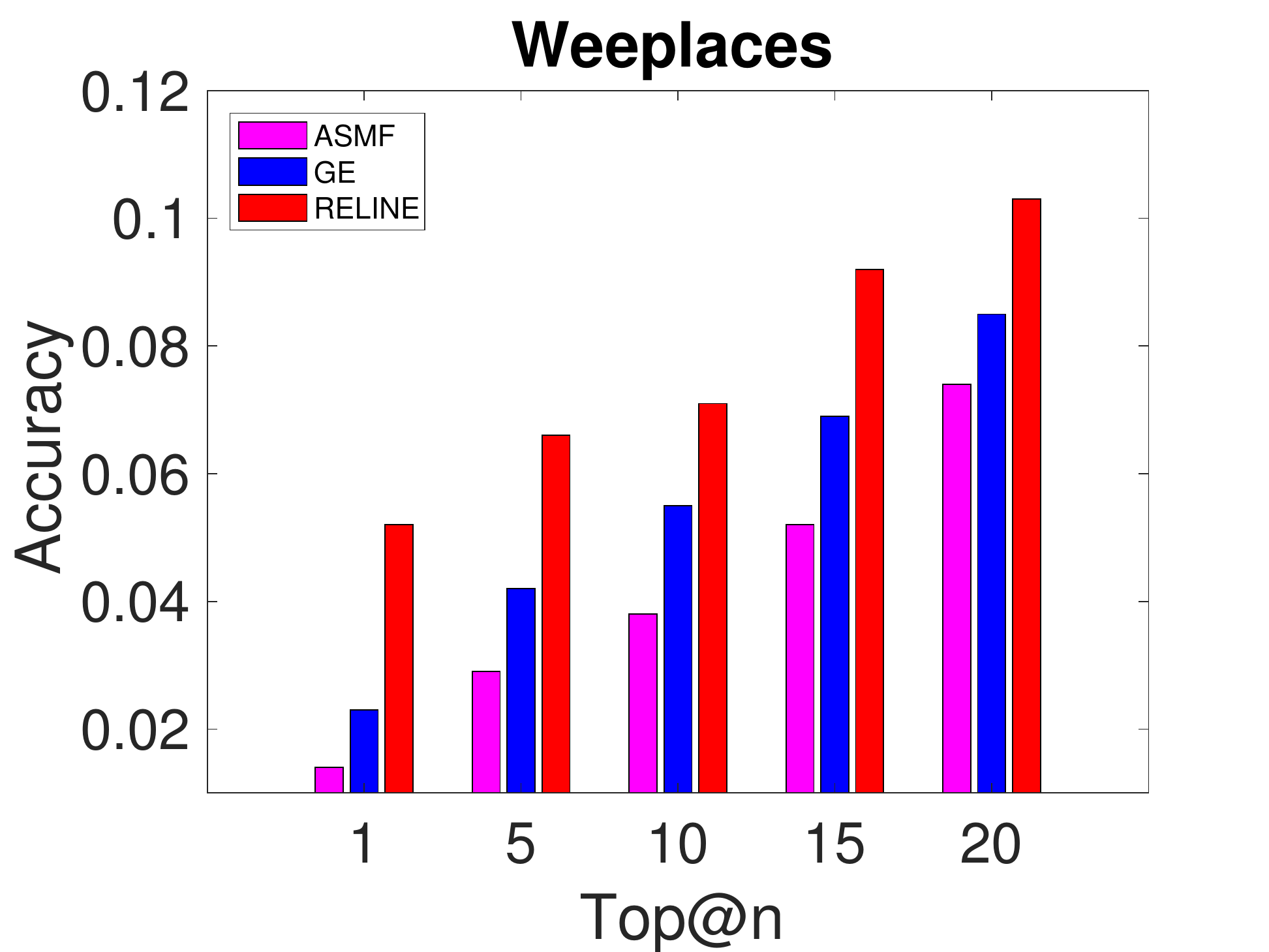}
		\label{fig:AccuracyWeeplacesCSP} 
		}
	\end{subfigure}	
	\begin{subfigure}[]{
	\centering	
\includegraphics[height=4.2cm, width=5.7cm]{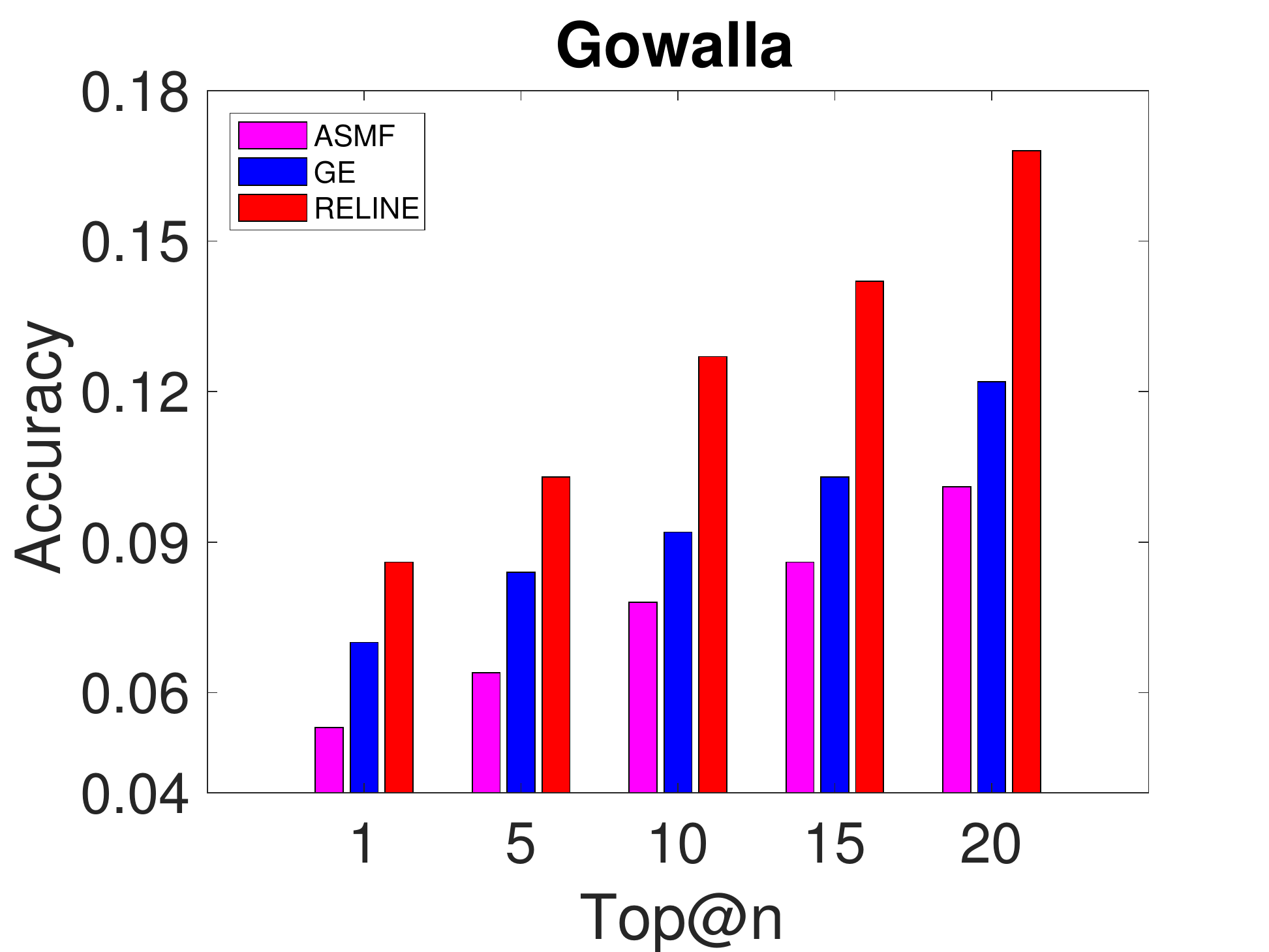}	
		\label{fig:AccuracyGowallaCSP}	  
		}
	\end{subfigure}
\vspace{-0.4cm}
\caption{Accuracy vs the top@$n$ recommendations for cold-start POIs.} 
\label{fig:DistributionsCSP} 
\end{figure*}

%% file: LaTeX/TunningParameters.tex
\begin{figure*}[!t]
\centering
	\begin{subfigure}[]{
	\centering			
\includegraphics[height=4cm, width=5.3cm]{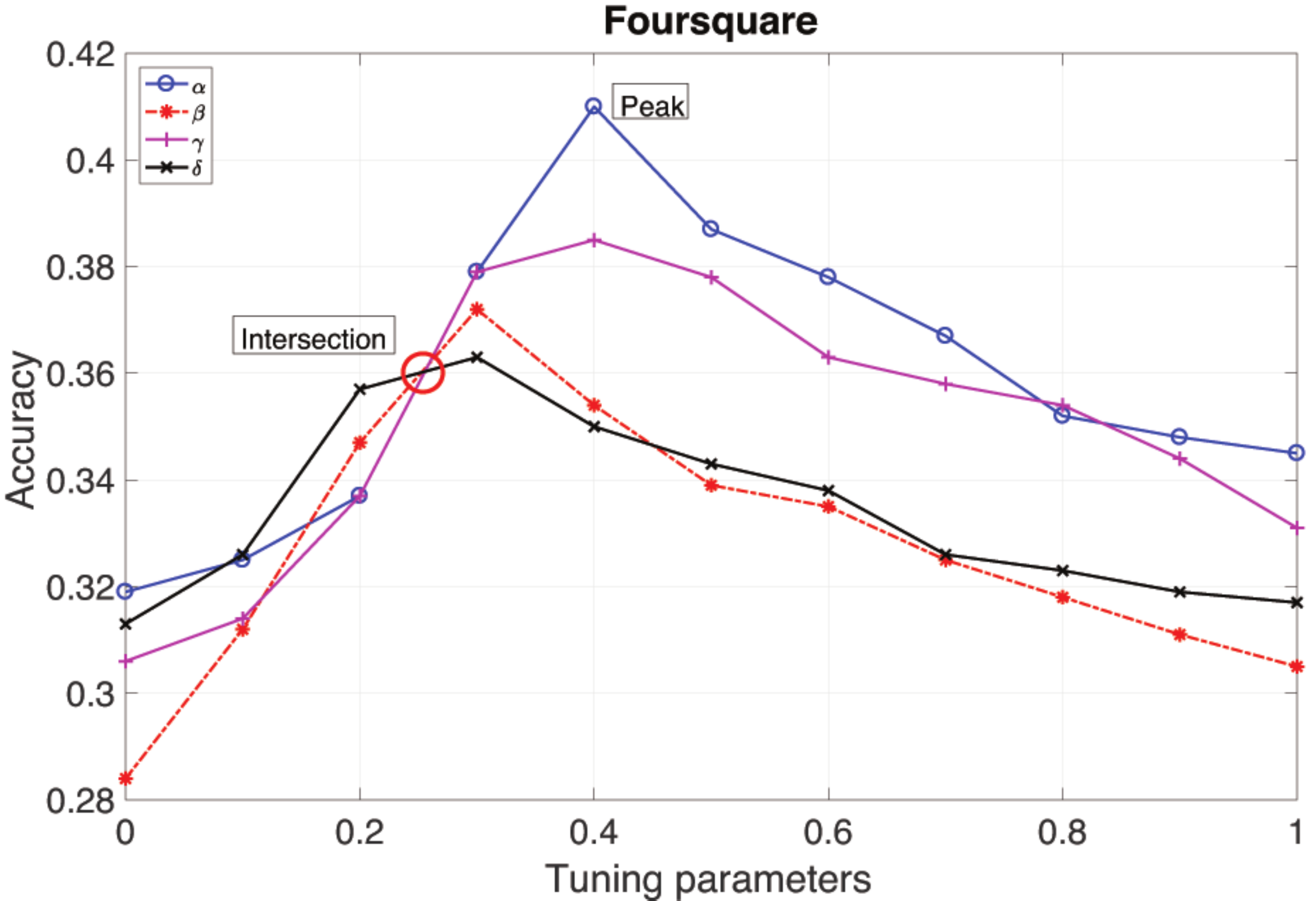}
		\label{fig:FoursquareParameters} 
		}
	\end{subfigure}
~	
	\begin{subfigure}[]{
	\centering			
\includegraphics[height=4cm, width=5.3cm]{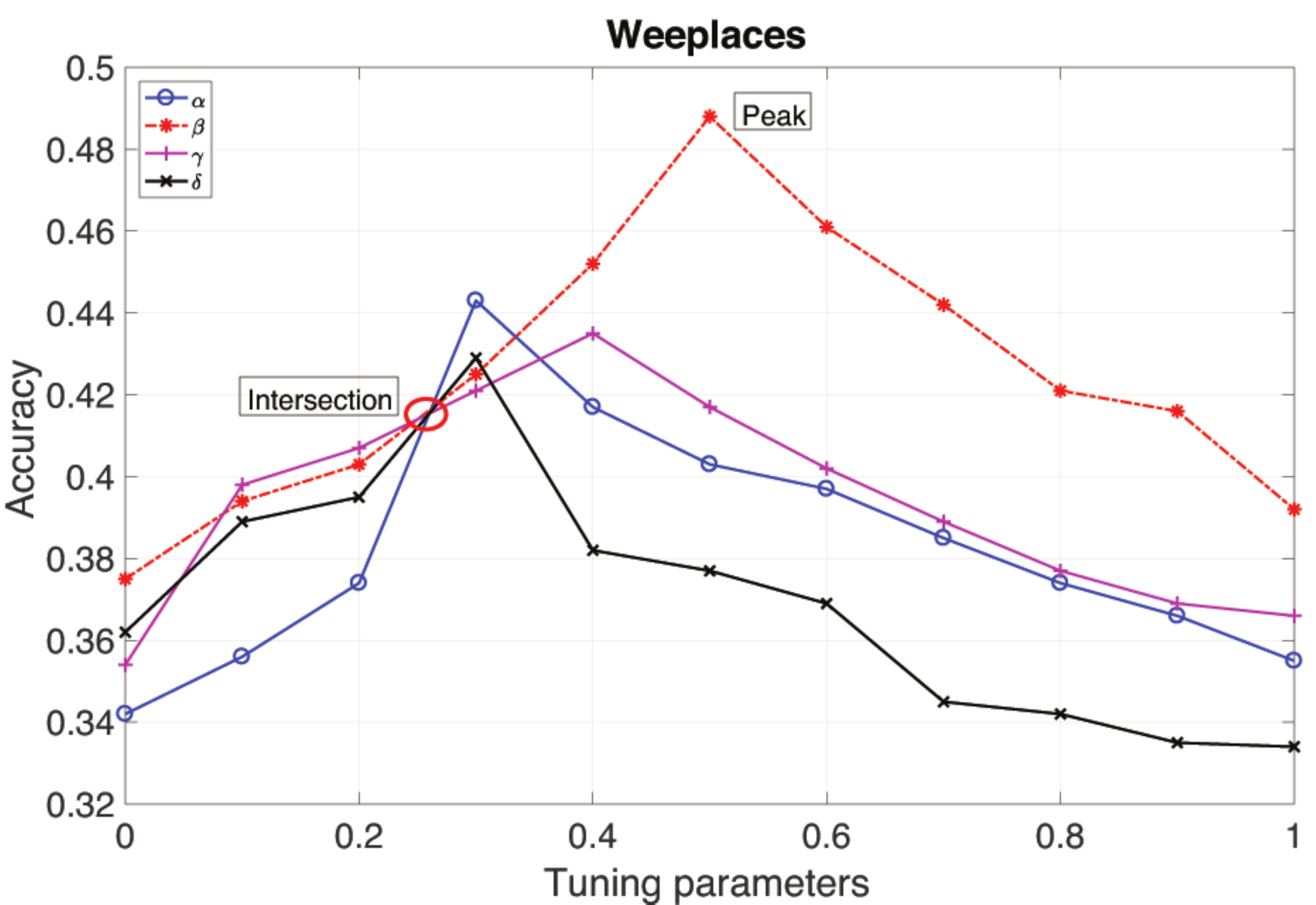}
		\label{fig:WeeplacesParameters} 
		}
	\end{subfigure}	
~
	\begin{subfigure}[]{
	\centering	
\includegraphics[height=4cm, width=5.3cm]{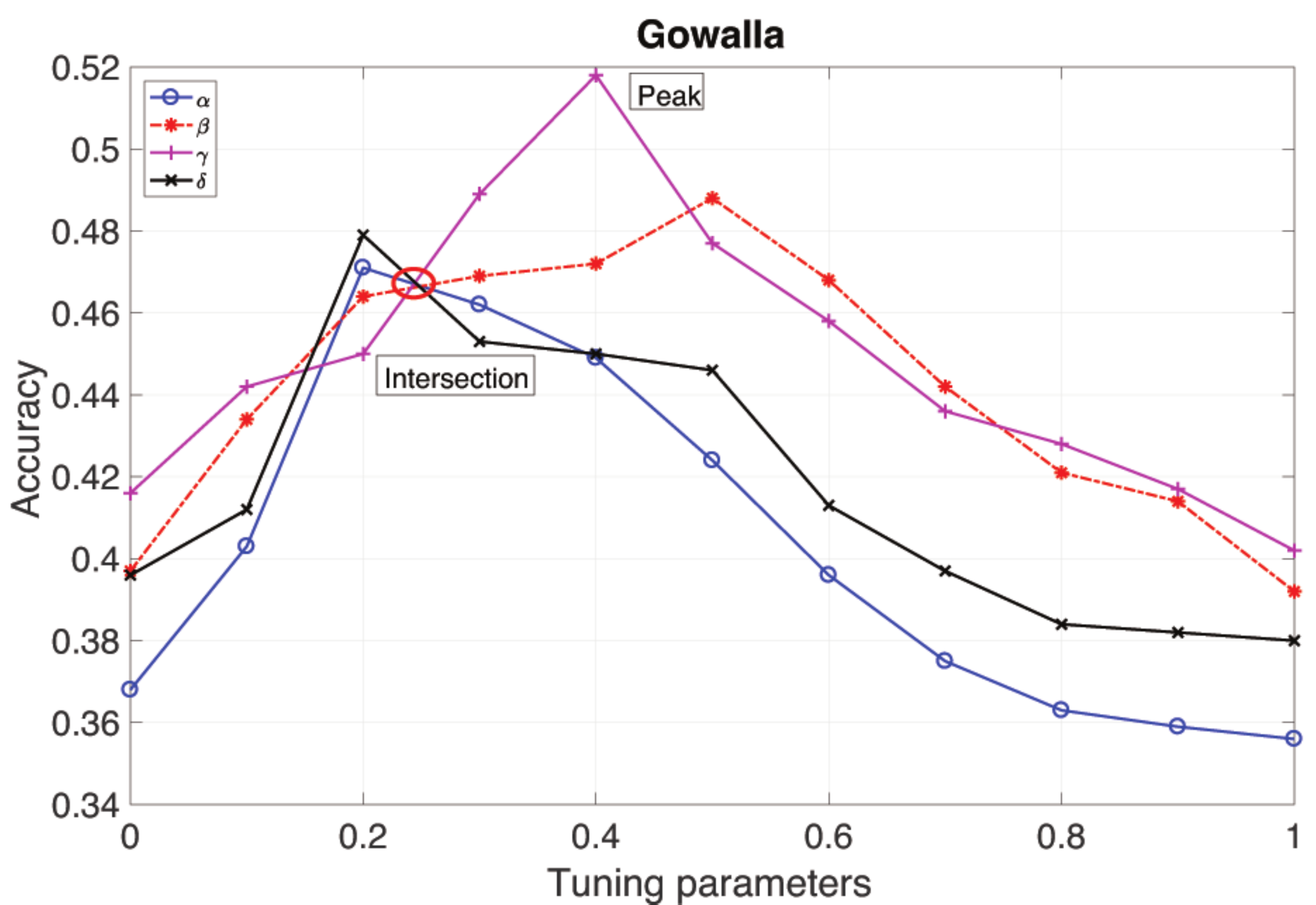}	
		\label{fig:GowallaParameters}	  
		}
	\end{subfigure}
\vspace{-0.4cm}
\caption{Tuning the parameters for all datasets.} 
\label{fig:Tuning parameters} 
\end{figure*}

%% file: conclusions.tex
\section{Conclusions}
\label{sec:Conclusion}

The rapid growth of users' interaction in OSN and the huge amount of information they provide, led researchers to models that retrieve personalized information close to their past preferences. Unfortunately, these models are tackled by sparsity and cold-start problems. The geographical information related to the content posted in such networks triggered new functionalities and directions for research to attack both problems. There are many models that provide POI recommendations using either the social network influence, or the geographical proximity of user's current location. However, these models miss the influence of the temporal dimension or users' preference dynamics, which are crucial while personalizing the retrieved information.

In this paper, we present a novel model that considers all aforementioned factors while providing POI recommendations. In particular, our model uses a probabilistic weighting strategy over 8 information graphs that correspond to users and POI relations. Then it uses a graph-based approach that jointly learns users' and POI embeddings from these weighted graphs into the same latent space and provides personalized POI recommendations. We examine the influence of social, geographical, temporal, and preference dynamics in terms of accuracy. We compare our approach against four state-of-the-art models measuring the accuracy of the recommendations for i) all the users, ii) the cold-start users, and iii) the cold-start locations. Our method significantly outperforms state-of-the-art approaches. 

%% file: main.bbl
\begin{thebibliography}{10}

\bibitem{Baral2016}
R.~Baral and T.~Li.
\newblock {MAPS}: A multi aspect personalized {POI} recommender system.
\newblock In {\em Proc. 10th ACM Conf. on Recommender Systems (RecSys)}, pages
  281--284, Boston, MA, 2016.

\bibitem{Cho2011}
E.~Cho, S.~A. Myers, and J.~Leskovec.
\newblock Friendship and mobility: User movement in location-based social
  networks.
\newblock In {\em Proc. 17th ACM SIGKDD International Conf. on Knowledge
  Discovery \& Data Mining (KDD)}, pages 1082--1090, San Diego, CA, 2011.

\bibitem{DSAA2018ckpm}
G.~Christoforidis, P.~Kefalas, A.~Papadopoulos, and Y.~Manolopoulos.
\newblock Recommendation of points-of-interest using graph embeddings.
\newblock In {\em Proc. 5th IEEE International Conf. on Data Science \&
  Advanced Analytics (DSAA)}, Turin, Italy, 2018.

\bibitem{Gao2015}
H.~Gao, J.~Tang, X.~Hu, and H.~Liu.
\newblock Content-aware point of interest recommendation on location-based
  social networks.
\newblock In {\em Proc. 29th AAAI Conf. on Artificial Intelligence}, pages
  1721--1727, Austin, TX, 2015.

\bibitem{Herlocker2004}
J.~L. Herlocker, J.~A. Konstan, L.~G. Terveen, and J.~T. Riedl.
\newblock Evaluating collaborative filtering recommender systems.
\newblock {\em ACM Transactions on Information Systems}, 22(1):5--53, 2004.

\bibitem{Kefalas2017}
P.~Kefalas and Y.~Manolopoulos.
\newblock A time-aware spatio-textual recommender system.
\newblock {\em Expert Systems with Applications}, 78:396--406, 2017.

\bibitem{Li2014}
A.~Q. Li, A.~Ahmed, S.~Ravi, and A.~J. Smola.
\newblock Reducing the sampling complexity of topic models.
\newblock In {\em Proc. 20th ACM SIGKDD International Conf. on Knowledge
  Discovery \& Data Mining (KDD)}, pages 891--900, New York, NY, 2014.

\bibitem{Li2016}
H.~Li, Y.~Ge, R.~Hong, and H.~Zhu.
\newblock Point-of-{I}nterest recommendations: Learning potential check-ins
  from friends.
\newblock In {\em Proc. 22nd ACM SIGKDD International Conf. on Knowledge
  Discovery \& Data Mining (KDD)}, pages 975--984, San Francisco, CA, 2016.

\bibitem{Li2015}
X.~Li, G.~Cong, X.-L. Li, T.-A.~N. Pham, and S.~Krishnaswamy.
\newblock {Rank-GeoFM}: A ranking based geographical factorization method for
  point of interest recommendation.
\newblock In {\em Proc. 38th International ACM Conf. on Research \& Development
  in Information Retrieval (SIGIR)}, pages 433--442, Santiago, Chile, 2015.

\bibitem{Liu2017}
B.~Liu, T.~Qian, B.~Liu, L.~Hong, Z.~You, and Y.~Li.
\newblock Learning spatiotemporal-aware representation for {POI}
  recommendation.
\newblock {\em CoRR}, abs/1704.08853, 2017.

\bibitem{Liu2014}
L.~Liu, J.~Xu, S.~S. Liao, and H.~Chen.
\newblock A real-time personalized route recommendation system for self-drive
  tourists based on vehicle to vehicle communication.
\newblock {\em Expert Systems with Applications}, 41(7):3409--3417, 2014.

\bibitem{Lu2010a}
Z.~Lu, B.~Savas, W.~Tang, and I.~S. Dhillon.
\newblock Supervised link prediction using multiple sources.
\newblock In {\em Proc. 10th IEEE International Conf. on Data Mining (ICDM)},
  pages 923--928, Sydney, Australia, 2010.

\bibitem{Meng2013}
F.~Meng, D.~Gao, W.~Li, X.~Sun, and Y.~Hou.
\newblock A unified graph model for personalized query-oriented reference paper
  recommendation.
\newblock {\em Proc. 22nd ACM International Conf. on Information \& Knowledge
  Management (CIKM)}, 2013.

\bibitem{Mikolov2013}
T.~Mikolov, I.~Sutskever, K.~Chen, G.~Corrado, and J.~Dean.
\newblock Distributed representations of words and phrases and their
  compositionality.
\newblock In {\em Proc. 26th International Conf. on Neural Information
  Processing Systems (NIPS)}, pages 3111--3119, Lake Tahoe, NV, 2013.

\bibitem{Niu2011}
F.~Niu, B.~Recht, C.~Re, and S.~J. Wright.
\newblock {HOGWILD!:} a lock-free approach to parallelizing stochastic gradient
  descent.
\newblock In {\em Proc. 24th International Conf. on Neural Information
  Processing Systems (NIPS)}, pages 693--701, Granada, Spain, 2011.

\bibitem{Tang2015}
J.~Tang, M.~Qu, M.~Wang, M.~Zhang, J.~Yan, and Q.~Mei.
\newblock {LINE}: Large-scale information network embedding.
\newblock In {\em Proc. of 24th International Conf. on World Wide Web (WWW)},
  pages 1067--1077, Florence, Italy, 2015.

\bibitem{Vasuki2010}
V.~Vasuki, N.~Natarajan, Z.~Lu, B.~Savas, and I.~Dhillon.
\newblock Scalable affiliation recommendation using auxiliary networks.
\newblock {\em ACM Transaction on Intelligent Systems \& Technology}, 3(1),
  2011.

\bibitem{Wang2017}
H.~Wang, Y.~Fu, Q.~Wang, H.~Yin, C.~Du, and H.~Xiong.
\newblock A location-sentiment-aware recommender system for both home-town and
  out-of-town users.
\newblock {\em Proc. 23rd ACM SIGKDD International Conf. on Knowledge Discovery
  \& Data Mining (KDD)}, 2017.

\bibitem{Wang2016}
W.~Wang, H.~Yin, S.~Sadiq, L.~Chen, M.~Xie, and X.~Zhou.
\newblock {SPORE}: A sequential personalized spatial item recommender system.
\newblock In {\em Proc. 32nd IEEE International Conf. on Data Engineering
  (ICDE)}, pages 954--965, Helsinki, Finland, 2016.

\bibitem{Xie2016}
M.~Xie, H.~Yin, H.~Wang, F.~Xu, W.~Chen, and S.~Wang.
\newblock Learning graph-based {POI} embedding for location-based
  recommendation.
\newblock In {\em Proc. 25th ACM International on Conf. on Information \&
  Knowledge Management (CIKM)}, pages 15--24, Indianapolis, IN, 2016.

\bibitem{Xiong2010}
L.~Xiong, X.~Chen, T.-K. Huang, J.~Schneider, and J.~G. Carbonell.
\newblock Temporal collaborative filtering with {B}ayesian probabilistic tensor
  factorization.
\newblock In {\em Proc. 10th SIAM International Conf. on Data Mining (SDM)},
  pages 211--222, 2010.

\bibitem{Yang2017}
C.~Yang, L.~Bai, C.~Zhang, Q.~Yuan, and J.~Han.
\newblock Bridging collaborative filtering and semi-supervised learning: A
  neural approach for {POI} recommendation.
\newblock In {\em Proc. 23rd ACM SIGKDD International Conf. on Knowledge
  Discovery \& Data Mining (KDD)}, pages 1245--1254, Halifax, Canada, 2017.

\bibitem{Yin2017}
H.~Yin, W.~Wang, H.~Wang, L.~Chen, and X.~Zhou.
\newblock Spatial-aware hierarchical collaborative deep learning for{POI}
  recommendation.
\newblock {\em IEEE Transactions on Knowledge and Data Engineering},
  29:2537--2551, 2017.

\bibitem{Yuan2013}
Q.~Yuan, G.~Cong, Z.~Ma, A.~Sun, and N.~M. Thalmann.
\newblock Time-aware point-of-interest recommendation.
\newblock In {\em Proc. 36th ACM International Conf. on Research \& Development
  in Information Retrieval (SIGIR)}, pages 363--372, Dublin, Ireland, 2013.

\bibitem{Zhang2014}
J.-D. Zhang, C.-Y. Chow, and Y.~Li.
\newblock {LORE}: Exploiting sequential influence for location recommendations.
\newblock In {\em Proc. 22nd ACM SIGSPATIAL International Conf. on Advances in
  Geographic Information System (GIS)}, pages 103--112, Dallas, TX, 2014.

\bibitem{Zhao2017}
S.~Zhao, T.~Zhao, I.~King, and M.~R. Lyu.
\newblock Geo-teaser.
\newblock {\em Companion Proc. 26th International Conf. on World Wide Web
  (WWW)}, pages 153--162, 2017.

\end{thebibliography}


\begin{thebibliography}{10}

\bibitem{Baral2016}
R.~Baral and T.~Li.
\newblock Maps: A multi aspect personalized poi recommender system.
\newblock In {\em Proceedings of the 10th ACM Conference on Recommender Systems
  (RecSys)}, pages 281--284, Boston, Massachusetts, USA, 2016.

\bibitem{Cao2010}
X.~Cao, G.~Cong, and C.~S. Jensen.
\newblock Mining significant semantic locations from {GPS} data.
\newblock {\em Proceedings VLDB}, 3(1-2):1009--1020, 2010.

\bibitem{Cho2011}
E.~Cho, S.~A. Myers, and J.~Leskovec.
\newblock Friendship and mobility: User movement in location-based social
  networks.
\newblock In {\em Proceedings of the 17th ACM SIGKDD International Conference
  on Knowledge Discovery and Data Mining}, pages 1082--1090, 2011.

\bibitem{Cremonesi2010}
P.~Cremonesi, Y.~Koren, and R.~Turrin.
\newblock Performance of recommender algorithms on top-n recommendation tasks.
\newblock In {\em Proceedings of the Fourth ACM Conference on Recommender
  Systems (RecSys)}, pages 39--46, 2010.

\bibitem{Gao2015}
H.~Gao, J.~Tang, X.~Hu, and H.~Liu.
\newblock Content-aware point of interest recommendation on location-based
  social networks.
\newblock In {\em Proceedings of the 29th AAAI Conference on Artificial
  Intelligence}, pages 1721--1727, 2015.

\bibitem{Herlocker2004}
J.~L. Herlocker, J.~A. Konstan, L.~G. Terveen, and J.~T. Riedl.
\newblock Evaluating collaborative filtering recommender systems.
\newblock {\em ACM Transactions on Information Systems (TIS)}, 22(1):5--53,
  2004.

\bibitem{Kefalas2017}
P.~Kefalas and Y.~Manolopoulos.
\newblock A time-aware spatio-textual recommender system.
\newblock {\em Expert Systems with Applications}, 78:396 -- 406, 2017.

\bibitem{Lathia2010}
N.~Lathia, S.~Hailes, L.~Capra, and X.~Amatriain.
\newblock Temporal diversity in recommender systems.
\newblock In {\em Proceedings 33rd International ACM Conference on Research \&
  Development in Information Retrieval (SIGIR)}, pages 210--217, 2010.

\bibitem{Lee2008}
T.~Q. Lee, Y.~Park, and Y.-T. Park.
\newblock A time-based approach to effective recommender systems using implicit
  feedback.
\newblock {\em Expert Systems with Applications}, 34(4):3055--3062, 2008.

\bibitem{Li2014}
A.~Q. Li, A.~Ahmed, S.~Ravi, and A.~J. Smola.
\newblock Reducing the sampling complexity of topic models.
\newblock In {\em Proceedings of the 20th ACM SIGKDD International Conference
  on Knowledge Discovery and Data Mining}, pages 891--900, 2014.

\bibitem{Li2016}
H.~Li, Y.~Ge, R.~Hong, and H.~Zhu.
\newblock Point-of-interest recommendations: Learning potential check-ins from
  friends.
\newblock In {\em Proceedings of the 22Nd ACM SIGKDD International Conference
  on Knowledge Discovery and Data Mining}, pages 975--984, 2016.

\bibitem{Li2015}
X.~Li, G.~Cong, X.-L. Li, T.-A.~N. Pham, and S.~Krishnaswamy.
\newblock Rank-geofm: A ranking based geographical factorization method for
  point of interest recommendation.
\newblock In {\em Proceedings of the 38th International ACM SIGIR Conference on
  Research and Development in Information Retrieval}, pages 433--442, 2015.

\bibitem{Liu2017}
B.~Liu, T.~Qian, B.~Liu, L.~Hong, Z.~You, and Y.~Li.
\newblock Learning spatiotemporal-aware representation for {POI}
  recommendation.
\newblock {\em CoRR}, abs/1704.08853, 2017.

\bibitem{Liu2014}
L.~Liu, J.~Xu, S.~S. Liao, and H.~Chen.
\newblock A real-time personalized route recommendation system for self-drive
  tourists based on vehicle to vehicle communication.
\newblock {\em Expert Systems with Applications}, 41(7):3409--3417, 2014.

\bibitem{Lu2010a}
Z.~Lu, B.~Savas, W.~Tang, and I.~S. Dhillon.
\newblock Supervised link prediction using multiple sources.
\newblock In {\em Proceedings 10th IEEE International Conference on Data Mining
  (ICDM)}, pages 923--928, 2010.

\bibitem{Meng2013}
F.~Meng, D.~Gao, W.~Li, X.~Sun, and Y.~Hou.
\newblock A unified graph model for personalized query-oriented reference paper
  recommendation.
\newblock {\em Proceedings of the 22nd ACM international conference on
  Conference on information and knowledge management (CIKM)}, 2013.

\bibitem{Mikolov2013}
T.~Mikolov, I.~Sutskever, K.~Chen, G.~Corrado, and J.~Dean.
\newblock Distributed representations of words and phrases and their
  compositionality.
\newblock In {\em Proceedings of the 26th International Conference on Neural
  Information Processing Systems (NIPS)}, pages 3111--3119, Lake Tahoe, Nevada,
  2013.

\bibitem{Niu2011}
F.~Niu, B.~Recht, C.~Re, and S.~J. Wright.
\newblock {HOGWILD!:} a lock-free approach to parallelizing stochastic gradient
  descent.
\newblock In {\em Proceedings of the 24th International Conference on Neural
  Information Processing Systems (NIPS)}, pages 693--701, 2011.

\bibitem{Sarwar2001}
B.~Sarwar, G.~Karypis, J.~Konstan, and J.~Riedl.
\newblock Item-based collaborative filtering recommendation algorithms.
\newblock In {\em Proceedings of the 10th International Conference on World
  Wide Web (WWW)}, pages 285--295, 2001.

\bibitem{Tang2015}
J.~Tang, M.~Qu, M.~Wang, M.~Zhang, J.~Yan, and Q.~Mei.
\newblock Line: Large-scale information network embedding.
\newblock In {\em Proceedings of the 24th International Conference on World
  Wide Web (WWW)}, pages 1067--1077, 2015.

\bibitem{Vasuki2010}
V.~Vasuki, N.~Natarajan, Z.~Lu, B.~Savas, and I.~Dhillon.
\newblock Scalable affiliation recommendation using auxiliary networks.
\newblock {\em ACM Transaction on Intelligent Systems \& Technology}, 3(1),
  2011.

\bibitem{Wang2017}
H.~Wang, Y.~Fu, Q.~Wang, H.~Yin, C.~Du, and H.~Xiong.
\newblock A location-sentiment-aware recommender system for both home-town and
  out-of-town users.
\newblock {\em Proceedings of the 23rd ACM SIGKDD International Conference on
  Knowledge Discovery and Data Mining}, 2017.

\bibitem{Wang2016}
W.~Wang, H.~Yin, S.~Sadiq, L.~Chen, M.~Xie, and X.~Zhou.
\newblock Spore: A sequential personalized spatial item recommender system.
\newblock In {\em 2016 IEEE 32nd International Conference on Data Engineering
  (ICDE)}, pages 954--965, Helsinki, Finland, 2016.

\bibitem{Xie2016}
M.~Xie, H.~Yin, H.~Wang, F.~Xu, W.~Chen, and S.~Wang.
\newblock Learning graph-based poi embedding for location-based recommendation.
\newblock In {\em Proceedings of the 25th ACM International on Conference on
  Information and Knowledge Management (CIKM)}, pages 15--24, 2016.

\bibitem{Xiong2010}
L.~Xiong, X.~Chen, T.-K. Huang, J.~Schneider, and J.~G. Carbonell.
\newblock Temporal collaborative filtering with bayesian probabilistic tensor
  factorization.
\newblock In {\em Proceedings 10th SIAM International Conference on Data Mining
  (SDM)}, pages 211--222, 2010.

\bibitem{Yang2017}
C.~Yang, L.~Bai, C.~Zhang, Q.~Yuan, and J.~Han.
\newblock Bridging collaborative filtering and semi-supervised learning: A
  neural approach for poi recommendation.
\newblock In {\em Proceedings of the 23rd ACM SIGKDD International Conference
  on Knowledge Discovery and Data Mining}, pages 1245--1254, 2017.

\bibitem{Yin2014}
H.~Yin, B.~Cui, Y.~Sun, Z.~Hu, and L.~Chen.
\newblock Lcars: A spatial item recommender system.
\newblock {\em ACM Transactions on Information Systems (TOIS)},
  32(3):11:1--11:37, July 2014.

\bibitem{Yin2017}
H.~Yin, W.~Wang, H.~Wang, L.~Chen, and X.~Zhou.
\newblock Spatial-aware hierarchical collaborative deep learning for poi
  recommendation.
\newblock {\em IEEE Transactions on Knowledge and Data Engineering (TKDE)},
  29:2537--2551, 2017.

\bibitem{Yin2015}
H.~Yin, X.~Zhou, Y.~Shao, H.~Wang, and S.~Sadiq.
\newblock Joint modeling of user check-in behaviors for point-of-interest
  recommendation.
\newblock In {\em Proceedings of the 24th ACM International on Conference on
  Information and Knowledge Management (CIKM)}, pages 1631--1640, 2015.

\bibitem{Yuan2013}
Q.~Yuan, G.~Cong, Z.~Ma, A.~Sun, and N.~M. Thalmann.
\newblock Time-aware point-of-interest recommendation.
\newblock In {\em Proceedings of the 36th ACM International Conference on
  Research \& Development in Information Retrieval (SIGIR)}, pages 363--372,
  2013.

\bibitem{Zhang2014}
J.-D. Zhang, C.-Y. Chow, and Y.~Li.
\newblock Lore: Exploiting sequential influence for location recommendations.
\newblock In {\em Proceedings of the 22Nd ACM SIGSPATIAL International
  Conference on Advances in Geographic Information Systems}, pages 103--112,
  2014.

\bibitem{Zhao2017}
S.~Zhao, T.~Zhao, I.~King, and M.~R. Lyu.
\newblock Geo-teaser.
\newblock {\em Proceedings of the 26th International Conference on World Wide
  Web Companion (WWW)}, 2017.

\end{thebibliography}
